\documentclass[12pt]{article}
\pdfoutput=1

\usepackage[T1]{fontenc}
\usepackage[bulletsep]{collref}
\usepackage{amssymb,graphicx}
\usepackage[intlimits]{amsmath}
\usepackage{bbm}
\usepackage{xcolor}
\usepackage[small]{subfigure}
\usepackage{wasysym}

\makeatletter \@addtoreset{equation}{section} \makeatother

\makeatletter
\let\old@startsection=\@startsection
\let\oldl@section=\l@section
\renewcommand{\@startsection}[6]{\old@startsection{#1}{#2}{#3}{#4}{#5}{#6\mathversion{bold}}}
\renewcommand{\l@section}[2]{\oldl@section{\mathversion{bold}#1}{#2}}
\makeatother

\makeatletter
\let\old@makecaption=\@makecaption
\def\@makecaption{\small\old@makecaption}
\makeatother

\renewcommand{\geq}{\geqslant}

\newcommand{\dd}{\mathbbm{D}}

\begin{document}

\thispagestyle{empty}
\begin{flushright}\footnotesize
\texttt{NORDITA 2020-109} 
\vspace{0.6cm}
\end{flushright}

\renewcommand{\thefootnote}{\fnsymbol{footnote}}
\setcounter{footnote}{0}

\begin{center}
{\Large\textbf{\mathversion{bold} Overlaps and Fermionic Dualities for  Integrable Super Spin Chains}
\par}

\vspace{0.8cm}

\textrm{Charlotte~Kristjansen$^{1}$, Dennis~M\"uller$^{1}$ and
Konstantin~Zarembo$^{1,2}$\footnote{Also at ITEP, Moscow, Russia}}
\vspace{4mm}

\textit{${}^1$Niels Bohr Institute, Copenhagen University, Blegdamsvej 17, 2100 Copenhagen, Denmark}\\
\textit{${}^2$Nordita, KTH Royal Institute of Technology and Stockholm University,
Roslagstullsbacken 23, SE-106 91 Stockholm, Sweden}\\
\vspace{0.2cm}
\texttt{kristjan@nbi.dk, dennis.muller@nbi.ku.dk, zarembo@nordita.org}

\vspace{3mm}


\par\vspace{1cm}

\textbf{Abstract} \vspace{3mm}

\begin{minipage}{13cm} 

The $\mathfrak{psu}(2,2|4)$ integrable super spin chain underlying the AdS/CFT correspondence has integrable boundary states which describe set-ups where $k$ D3-branes get
dissolved in a probe D5-brane. Overlaps between Bethe eigenstates and these boundary states encode the one-point functions of conformal operators and are expressed in terms of the superdeterminant of the Gaudin matrix that in turn depends on the Dynkin diagram of the symmetry algebra.  The different possible Dynkin diagrams of super Lie algebras are related via fermionic dualities and we determine how overlap formulae transform under these dualities. As an application we show
how to consistently move between overlap formulae obtained for $k=1$  from different Dynkin diagrams.

\end{minipage}
\end{center}

\vspace{0.5cm}


\setcounter{page}{1}
\renewcommand{\thefootnote}{\arabic{footnote}}
\setcounter{footnote}{0}


\section{Introduction}

 The study of the AdS/CFT correspondence with the presence of defects has lead to the discovery of a number of integrable boundary states of the $\mathfrak{psu}(2,2|4)$  super spin chain underlying ${\cal N}=4$ SYM, among these certain matrix product and valence bond states~\cite{deLeeuw:2015hxa,Buhl-Mortensen:2015gfd,Komatsu:2020sup,Gombor:2020kgu,Kristjansen:2020mhn,Gombor:2020auk}. The overlap between these particular boundary states and the Bethe eigenstates of the 
spin chain encode information about the one-point functions of conformal operators in  domain wall versions of ${\cal N}=4$ SYM~\cite{deLeeuw:2015hxa,Buhl-Mortensen:2015gfd}, and other overlaps are of relevance for the study of quantum quenches in statistical mechanics~\cite{Piroli:2017sei}.
  Not least the AdS/dCFT motivation has sparked the derivation of a number of exact overlap formulae. 
  From the first exact  expression  involving the overlap between the Bethe eigenstates of the Heisenberg spin chain and the N\'{e}el state, obtained in statistical physics~\cite{Pozsgay:2009,Brockmann:2014a,Brockmann:2014b},  
the catalogue of exact formulae has been extended to overlaps with a large class of matrix product states~\cite{deLeeuw:2015hxa,Buhl-Mortensen:2015gfd} and arbitrary valence bond states ~\cite{Pozsgay:2018ixm}, as well as to overlaps in several  bosonic spin chains where nesting is involved~\cite{deLeeuw:2016umh,deLeeuw:2018mkd,Piroli:2018ksf,Piroli:2018don,deLeeuw:2019ebw}.  The latest addition consists of overlap formulae for integrable super spin chains~\cite{Komatsu:2020sup,Gombor:2020kgu,Kristjansen:2020mhn,Gombor:2020auk}. 

 All known overlap formulae contain as a key ingredient the Gaudin matrix~\cite{Gaudin:1976sv}  of the Bethe eigenstate, or more precisely an object which can be expressed as the superdeterminant of  the Gaudin matrix~\cite{Kristjansen:2020mhn}.  The Gaudin matrix encodes the norm of the Bethe 
eigenstate~\cite{Gaudin:1976sv,Korepin:1982gg}  and can be expressed in a closed form given the Bethe roots of the state plus the Cartan matrix and Dynkin labels describing the Lie algebra and its particular representation underlying the integrable spin chain in question.  Dynkin diagrams and Cartan matrices for super Lie algebras are not unique
 \cite{Frappat:1996pb} but related via a set of fermionic dualities \cite{Tsuboi:1998ne}, 
 and this immediately raises a question in relation to the newly derived overlap formulae for integrable super spin chains, namely: How do these formulae transform under fermionic dualities? This question is the main focus of our work. We argue that for consistency reasons the overlap formulae have to transform
covariantly under fermionic dualities, a property to be defined more precisely in the following, and that this requirement puts very strong constraints on these formulae.  Furthermore, we  derive the transformation properties of the super determinant of the Gaudin matrix for all fermionic dualities that are needed to move between the various possible
Dynkin diagrams of $\mathfrak{psu}(2,2|4)$. As an application we transform an overlap formula found in~\cite{Gombor:2020auk} for the  Dynkin diagram corresponding to the
alternating grading to the cases of the Beauty and the Beast Dynkin diagram~\cite{Beisert:2005fw}. Transformation rules that we derive permit us to check the general formula for one-point functions~\cite{Gombor:2020auk} against explicit field-theory calculations presented in~\cite{Kristjansen:2020mhn} in a different grading.

Our paper is organized as follows.  We begin in section~\ref{Integrable_overlaps} by describing the general structure of overlap formulae for integrable boundary states where the 
superdeterminant of the Gaudin matrix plays a key role.
We furthermore review how fermionic dualities allow one to move between different Dynkin diagrams and associated Cartan matrices of a super Lie algebra and correspondingly between different sets of Bethe equations
determining the eigenstates of  the integrable super spin chain in question.  Then, in section~\ref{sec:TransLaws}, we determine how the superdeterminant of the Gaudin matrix transforms under fermionic dualities treating first the dualization after a non-momentum-carrying node and subsequently the slightly more complicated case of dualization after a momentum-carrying node where the superdeterminant  becomes singular and needs regularization. In both cases we start by a simple example and work our way towards the general case.  With the transformation rules for the superdeterminant in place  we, in
section~\ref{sec:DualizingOF}, turn to 
the translation of overlap formulae between different gradings starting by going through the procedure in some detail for
$\mathfrak{su}(2|2)$ and finally demonstrating how to translate the overlap formulae of $\mathfrak{psu}(2,2|4)$ between any
two gradings.  Section~\ref{conclusion} contains our conclusion.

\section{Integrable Overlaps and Fermionic Duality \label{Integrable_overlaps}}

\subsection{Overlap Formulae}

The Bethe-ansatz spectrum of an integrable spin chain with a rational R-matrix is neatly encoded in the group-theory data. Each  eigenstate is characterized by the rapidities of constituent magnons $u_{ja}$ assigned to the nodes of the Dynkin diagram. The spectral equations depend on the Cartan matrix  $M_{ab}$ and the Dynkin labels of the spin representation $q_a$:
\begin{equation}\label{BAEs}
 \left(\frac{u_{aj}-\frac{iq_a}{2}}{u_{aj}+\frac{iq_a}{2}}\right)^L
 \prod_{bk}\frac{u_{aj}-u_{bk}+\frac{iM_{ab}}{2}}{u_{aj}-u_{bk}-\frac{iM_{ab}}{2}}\equiv \,{\rm e}\,^{i\chi _{aj}}=-1.
\end{equation}
Their solutions  enumerate all eigenstates of the Hamiltonian.

Local single-trace operators in the $\mathcal{N}=4$ SYM correspond to 
the Bethe states $\left|\left\{u_{aj}\right\}\right\rangle\equiv \left|\mathbf{u}\right\rangle$ of the $\mathfrak{psu}(2,2|4)$ spin chain  
\cite{Beisert:2003jj}, equivalently represented via the AdS/CFT duality by on-shell states of the dual string theory. Likewise, the boundary states of the spin chain describe D-branes, which take variety of forms in gauge theory. Expectation values of local operators induced by the D-brane are naturally given by an overlap between the boundary state and on-shell Bethe eigenstates. This description proved highly efficient in computations of correlation functions in the presence of domain walls \cite{deLeeuw:2015hxa,Buhl-Mortensen:2015gfd} or of very large determinant operators \cite{Jiang:2019xdz,Jiang:2019zig}.  

An example is the D3-D5 domain wall, a codimension one defect across which the $SU(N)$ symmetry of the SYM is broken to $SU(N-1)$. We shall consider the case where symmetry breaking is effected by Neumann/Dirichlet boundary conditions imposed on the sundry field components \cite{Gaiotto:2008sa}. The D3-D5 defect preserves scaling symmetry, but allows for non-trivial one-point functions with power-law fall-off. An expectation value of the local operator $\mathcal{O}_\mathbf{u}$ in the presence of the domain wall at $x_3=0$ is thus given by 
\begin{equation}\label{1pt}
 \left\langle \mathcal{O}_{\mathbf{u}}(x)\right\rangle=
  \frac{\left\langle {\rm D3D5}\right.\!\left| \mathbf{u}\right\rangle}{\left\langle \mathbf{u}\right.\!\left|\mathbf{u} \right\rangle^{\frac{1}{2}}}\,\,
  \frac{2^{1-L}L^{-\frac{1}{2}}}{x_3^L}\,,
\end{equation}
where the bracket $\left\langle {\rm D3D5}\right|$ denotes a boundary state in the $\mathfrak{psu}(2,2|4)$ spin chain.  It can be explicitly constructed  in perturbation theory by evaluating Feynman diagrams in the presence of the 
defect \cite{Buhl-Mortensen:2016pxs,Buhl-Mortensen:2016jqo,Buhl-Mortensen:2017ind,Kristjansen:2020mhn}. The combinatorial prefactor that depends on the length is just a matter of convention. 

A non-perturbative solution for the overlap was obtained by bootstrapping scattering theory of magnons off the D5-brane \cite{Komatsu:2020sup,Gombor:2020kgu,Gombor:2020auk}, with no reference to the explicit form of the  boundary wavefunction.
This was possible because the D3-D5 system preserves integrability and scattering off the D5-brane is completely elastic. 

By definition, an integrable boundary state is a coherent superposition of  magnon pairs with opposite momenta \cite{Ghoshal:1993tm,Piroli:2017sei}, and has non-zero projections only on parity-invariant Bethe eigenstates. We will assume that parity uniformly flips all the rapidities  $u_{aj}\rightarrow -u_{aj}$, as it does in the D3-D5 case, but more generally it can also permute nodes of the Dynkin diagram as exemplified in \cite{Jiang:2019xdz}. The Bethe roots in a parity-even state are either paired: $\left\{u_{aj},-u_{aj}\right\}$, $j=1,..,K_a/2$, or lie exactly at zero. We denote levels with zero roots by $a_\alpha $, $\alpha =1,\ldots ,\nu $. 

Another way to characterize a Bethe state is by the Q-functions 
\begin{equation}
 \mathcal{Q}_a(u)=\prod_{j=1}^{\mathcal{K}_a}(u-u_{aj}),
\end{equation}
in terms of which the Bethe equations (\ref{BAEs}) can be concisely written as
\begin{equation}\label{QBAEs}
 \frac{Q_\theta^{[-q_a]} (u_{aj})}{Q_\theta^{[+q_a]} (u_{aj})}\,
 \prod_{b}^{}\frac{\mathcal{Q}_{b}^{[+M_{ab}]}(u_{aj})}{\mathcal{Q}_{b}^{[-M_{ab}]}(u_{aj})}=-1,
\end{equation}
where
\begin{equation}
 Q_\theta (u)=u^L,
\end{equation}
and we use the standard notations
\begin{equation}
 f^\pm(u)=f\left(u\pm\frac{i}{2}\right),\qquad 
 f^{[\pm q]}(u)=f\left(u\pm\frac{iq}{2}\right).
\end{equation}

The states we consider have definite parity, and their Q-functions are either even or odd.  It will prove useful to deal which the reduced Baxter functions that are uniformly even:
\begin{equation}
 Q_a(u)=\prod_{j=1}^{\frac{K_a}{2}}\left(u^2-u_{ja}^2\right),
  \label{eqn:ReducedBaxter}
\end{equation}
where $K_a$ is the number of non-zero roots. Thus, $\mathcal{Q}_a(u)=Q_a(u)$ if the $a$-th node does not contain a root at zero, and  $\mathcal{Q}_a(u)=uQ_a(u)$ if it does.

\begin{figure}[t]
\begin{center}
 \centerline{\includegraphics[width=7cm]{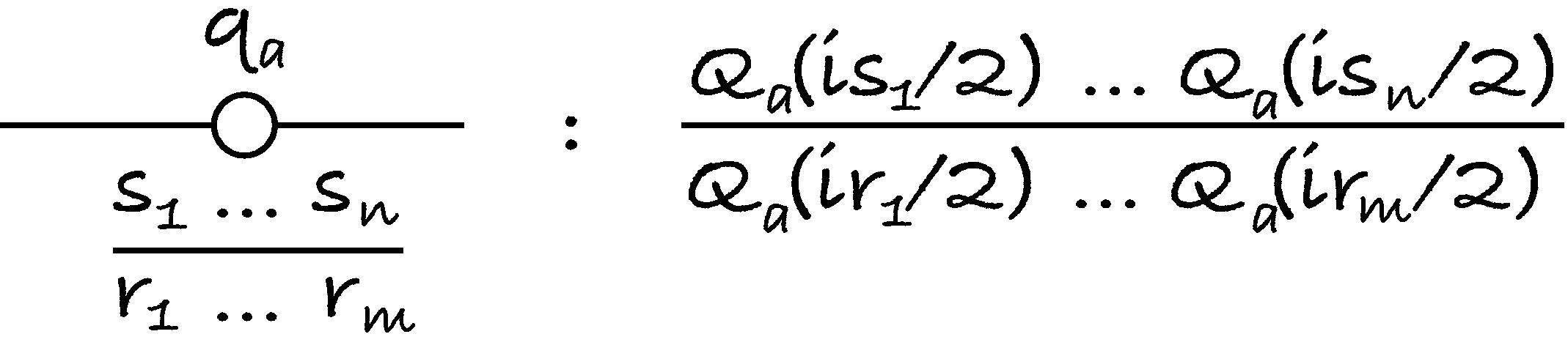}}
\caption{\label{gnot}\small A graphic notation for the pre-factor in the overlap formula.}
\end{center}
\end{figure}

The main ingredient of the overlap formulae is the Gaudin matrix, the Jacobian of the transformation from rapidities to phases in the Bethe equations:
\begin{equation}
 G_{aj,bk}=\frac{\partial \chi _{aj}}{\partial u_{bk}}\,.
\end{equation}
Parity is a linear $\mathbbm{Z}_2$ automorphism on the space of Bethe roots and this automorphism can be used to define the superdeterminant 
\begin{equation}
 \label{1GaudinSuper}
 \dd=\mathop{\mathrm{Sdet}}G.
\end{equation}
For any integrable boundary state known so far the overlap with on-shell Bethe states admits a determinant representation:
\begin{equation}
 \frac{\left\langle B\right.\!\left|\mathbf{u} \right\rangle^2}{\left\langle \mathbf{u}\right.\!\left|\mathbf{u} \right\rangle}
 =\prod_{a}^{}\frac{\prod\limits_{i=1}^{n}Q_{a}\left(\frac{is_{ai}}{2}\right)}{\prod\limits_{j=1}^{m}Q_a\left(\frac{ir_{aj}}{2}\right)}\,\dd,
\end{equation}
or an overlap may be given by a linear combination of  such terms. In view of its universal significance   we introduce for this formula a graphic notation shown  in fig.~\ref{gnot}.

Diagonalizing the $\mathbbm{Z}_2$ symmetry brings the Gaudin matrix into a block-diagonal form \cite{Brockmann:2014a}, and its superdeterminant can be expressed as a ratio of ordinary determinants: 
\begin{equation}
 \label{GaudinSuper}
 \dd=\frac{\det G^+}{\det G^-}\,,
\end{equation}
where $G^+$ and $G^-$ are $(K/2+\nu)\times  (K/2+\nu)$ and $K/2\times K/2$ matrices, respectively, with the matrix elements 
\begin{eqnarray}
  G^\pm_{aj,bk}&=&\left(\frac{Lq_a}{u_{aj}^2+\frac{q_a^2}{4}}-\sum_{cl}^{}K^+_{aj,cl}
 -\frac{1}{2}\sum_{\alpha }^{}\,K^+_{aj,a_\alpha 0}\right)\delta _{ab}\delta _{jk}+K^\pm_{aj,bk} \, ,
\nonumber \\
 G^+_{aj,\alpha }&=&\frac{1}{\sqrt{2}}\,K^+_{aj,a_\alpha 0}=G^+_{\alpha ,aj} \, ,
\nonumber \\
G^+_{\alpha \beta }&=&\left(\frac{4L}{q_{a_\alpha }}-\sum_{cl}K^+_{a_\alpha 0,cl}-\sum_{\gamma }^{}\frac{4}{M_{a_\alpha a_\gamma }}\right)
 \delta _{\alpha \beta }+\frac{4}{M_{a_\alpha b_\beta }}\,,
\end{eqnarray}
and
\begin{equation}
 K_{aj,bk}^\pm=\frac{M_{ab}}{(u_{aj}-u_{bk})^2+\frac{M_{ab}^2}{4}}\pm\frac{M_{ab}}{(u_{aj}+u_{bk})^2+\frac{M_{ab}^2}{4}}\,.
\end{equation}
If $q_a=0$, $1/q_\alpha $ should be set to zero, and the same for $M_{ab}$.

\begin{figure}[t]
\begin{center}
 \centerline{\includegraphics[width=10cm]{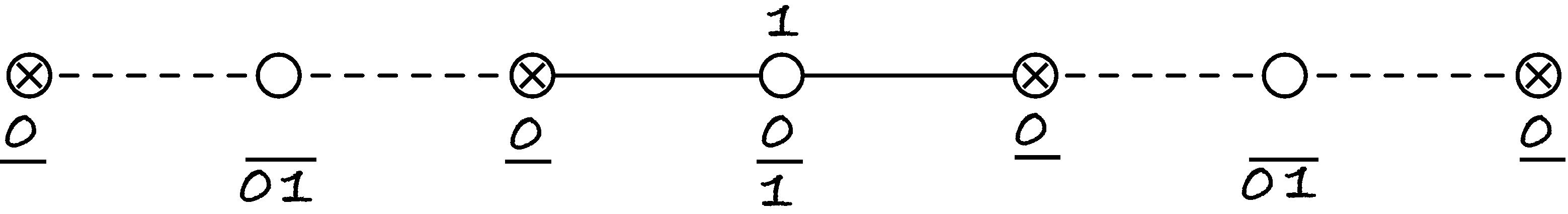}}
\caption{\label{Dynkin-Alt}\small Dynkin diagram of $\mathfrak{psu}(2,2|4)$ in the alternating grading. Numbers in the lower row represent Baxter functions in the overlap formula, according to the conventions of fig.~\ref{gnot}.
}
\end{center}
\end{figure}

The overlaps of the  D3-D5 boundary state (\ref{1pt}) were contructed in \cite{Gombor:2020kgu,Gombor:2020auk} from the non-perturbative solution of the boundary bootstrap \cite{Komatsu:2020sup,Gombor:2020kgu,Gombor:2020auk}. The weak-coupling limit of the asymptotic overlap formula reads\footnote{To take the weak-coupling limit, we set $s=x_s=1/2$ in (139) of \cite{Gombor:2020kgu}, rescale $w_j^{(1)}=u_{2j}/g$, $w_j^{(2)}=u_{6j}$, redefine $y^{(1)}_j=u_{3j}/g$ or $y^{(1)}_j=g/u_{1j}$, and $y^{(2)}_j=u_{5j}/g$ or $y^{(2)}_j=g/u_{7j}$, and send $g\rightarrow 0$. Here $g=\sqrt{\lambda }/4\pi $ is the gauge coupling. Two ways to scale the $y_j^{(1,2)}$ roots follow from the dynamic symmetry of the asymptotic Bethe equations \cite{Beisert:2005fw}. The asymptotic overlap formula holds under assumption of parity invariance of the Bethe state, inhereted by the perturbative overlap we are going to study.}:
\begin{equation}\label{D3D5-Alt}
 \frac{\left\langle {\rm D3D5}\right.\!\left| \mathbf{u}\right\rangle^2}{\left\langle \mathbf{u}\right.\!\left|\mathbf{u} \right\rangle}
 =\frac{Q_1(0)Q_3(0)Q_4(0)Q_5(0)Q_7(0)}{Q_2(0)Q_2\left(\frac{i}{2}\right)Q_4\left(\frac{i}{2}\right)Q_6(0)Q_6\left(\frac{i}{2}\right)}\,\mathbbm{D}.
\end{equation}
This formula corresponds to the Bethe equations in the alternating grading of the $\mathfrak{psu}(2,2|4)$ Dynkin diagram:
\begin{equation}
 M=\begin{bmatrix}
  0 & 1 &  &  &  &  &  \\ 
  1 & -2 & 1 &  &  &  &  \\ 
   & 1 & 0 & -1 &  &  &  \\ 
   &  & -1 & 2 & -1  &  &  \\ 
   &  &  &  -1 & 0 & 1 &  \\ 
   &  &  &  & 1 & -2 & 1 \\ 
   &  &  &  &  & 1 & 0 \\ 
 \end{bmatrix},
\end{equation}
and is illustrated in fig.~\ref{Dynkin-Alt}.

It would be interesting to compare this formula, derived by bootstrapping the boundary reflection, with the direct field-theory computations. The latter are available in a number of cases, but are presented in different gradings  \cite{deLeeuw:2018mkd,Kristjansen:2020mhn}  making direct comparison impossible beyond the simplest $\mathfrak{su}(2)$ sector. The Bethe equations (at one loop) can be transformed to any grading by a chain of fermionic dualities, which we review below, and one may expect that the overlap formulas make sense in any grading as well. Understanding how overlaps transform under fermionic duality is the main goal of this paper.

\subsection{Fermionic Duality}

Consider a fragment of the Cartan matrix and of the weight vector around an auxiliary, non-momentum-carrying fermionic node:
\begin{equation}\label{fragment}
 M=\begin{bmatrix}
 \ddots & 1  &  \\ 
  1 & 0 & -1 \\ 
  & -1 & \ddots\\ 
 \end{bmatrix}\qquad 
 q=\begin{bmatrix}
  \vdots \\ 
  0 \\ 
  \vdots \\ 
 \end{bmatrix}.
\end{equation}
The Bethe equations for the fermionic roots are expressed in terms of the Q-functions on the adjacent nodes:
\begin{equation}
 1=\frac{\mathcal{Q}_l^+(u_j)\mathcal{Q}_r^-(u_j)}{\mathcal{Q}_l^-(u_j)\mathcal{Q}_r^+(u_j)} \, ,
\end{equation}
where $\mathcal{Q}_l$, $\mathcal{Q}_r$ correspond to the upper and lower row in (\ref{fragment}), i.e.\ to the left and
right neigbouring node in the Dynkin diagram.

The fermionic duality is expressed by the equation \cite{Tsuboi:1998ne}\footnote{For an introduction to QQ-relations
and the ensuing fermionic and bosonic dualities we refer to~\cite{Gromov:2017blm}.}:
\begin{equation}\label{basicFD}
 \mathcal{Q}_l^-\mathcal{Q}_r^+-\mathcal{Q}_l^+\mathcal{Q}_r^-
 =i(\mathcal{K}_r-\mathcal{K}_l)\mathcal{Q}\widetilde{\mathcal{Q}},
\end{equation}
where $\mathcal{Q}$ is the Baxter polynomial on the fermionic node (middle row of (\ref{fragment})) and $\widetilde{\mathcal{Q}}$ is a new, dual Q-function. It is easy to see that  the original Bethe equations and the $QQ$-relation are equivalent to one another, because the left-hand side of the latter evaluates to zero on any root of $\mathcal{Q}(u)$. The dual polynomial $\widetilde{\mathcal{Q}}(u)$, of degree $\mathcal{K}_l+\mathcal{K}_r-\mathcal{K}-1$, absorbs the ``unused'' roots\footnote{Obviously, the Bethe equations have no solutions with $\mathcal{K}>\mathcal{K}_l+\mathcal{K}_r-1$.
The condition $\mathcal{K}_l+\mathcal{K}_r-\mathcal{K}>0$ imposes constraints on the quantum numbers of admissible states, typically equivalent to the highest-weight conditions. For example,  the number of roots should decrease from the middle to the wings for the Bethe equations labelled by the Dynkin diagram in fig.~\ref{Dynkin-Alt} \cite{Beisert:2003yb}.}. Quite obviously, the dual roots satisfy the same set of Bethe equations.

The fermionic Q-function enters the bosonic Bethe equations through the factor $\mathcal{Q}^+/\mathcal{Q}^-$  (cf.~(\ref{QBAEs})) that needs to be re-expressed  through the dual roots, if we want to completely dualize the full set of Bethe equations. The duality equation suffices to do so. Indeed, setting $u=u_{rk}\pm i/2$ consequtively in (\ref{basicFD}), and taking the ratio of the resulting identities gives:
\begin{equation}
 \frac{\mathcal{Q}^+(u_{lk})}{\mathcal{Q}^-(u_{lk})}
 =-\frac{\widetilde{\mathcal{Q}}^-(u_{lk})\mathcal{Q}^{++}_l(u_{lk})}{\widetilde{\mathcal{Q}}^+(u_{lk})\mathcal{Q}^{--}_l(u_{lk})}\,,
\end{equation}
and, likewise,
\begin{equation}\label{right-phase}
 \frac{\mathcal{Q}^-(u_{rk})}{\mathcal{Q}^+(u_{rk})}
 =-\frac{\widetilde{\mathcal{Q}}^+(u_{rk})\mathcal{Q}^{--}_r(u_{rk})}{\widetilde{\mathcal{Q}}^-(u_{rk})\mathcal{Q}^{++}_r(u_{rk})}\,.
\end{equation}
These are precisely the fermion factors in the Bethe equations to the left and to the right of the fermionic node. In addition to the dualized Q-functions they contain an extra self-scattering term that either cancels or reintroduces interactions among the $u_{lj}$ and $u_{rj}$ roots.

It is a easy to see that this extra factor reverses the grading of the two nodes at hand. Imagine the right node were bosonic. Its Bethe equations then contained the ratio $\mathcal{Q}_r^{++}/\mathcal{Q}_r^{--}$, exactly inverse to the Q-functions emerging from the duality transformation. Cancellation renders the node fermionic. If the node were fermionic from the beginning, the self-scattering induced by the duality makes the node bosonic with the correct phase. All in all, the Cartan matrix transforms as
\begin{equation}
  \begin{bmatrix}
\left\{ \genfrac{}{}{0pt}{}{-2}{0} \right\}  & 1  & 0 \\ 
 1  & 0 & -1 \\ 
  0 & -1  &  \left\{\genfrac{}{}{0pt}{}{2}{0} \right\} \\ 
 \end{bmatrix}~\longrightarrow~
 \begin{bmatrix}
 \left\{\genfrac{}{}{0pt}{}{0}{2} \right\}  & -1  & 0 \\ 
 -1  & 0 & 1 \\ 
  0 & 1  &  \left\{\genfrac{}{}{0pt}{}{0}{-2} \right\} \\ 
 \end{bmatrix}.
\end{equation}

The duality equation is slightly different if the fermionic node is momen\-tum-carrying \cite{Tsuboi:1998ne}:
\begin{equation}
 Q_\theta ^{[+q]}\mathcal{Q}_l^-\mathcal{Q}_r^+-Q_\theta ^{[-q]}\mathcal{Q}_l^+\mathcal{Q}_r^-
 =i(qL+\mathcal{K}_r-\mathcal{K}_l)\mathcal{Q}\widetilde{\mathcal{Q}},
\end{equation}
where  $q$ is the Dynkin label of the fermionic node:
\begin{equation}
 q=\begin{bmatrix}
  q_l \\ 
  q \\ 
  q_r \\ 
 \end{bmatrix}.
\end{equation}

The same manipulations as above now introduce an extra momentum phase:
\begin{eqnarray}
\frac{\mathcal{Q}^+(u_{lk})}{\mathcal{Q}^-(u_{lk})}
 &=&-\frac{\widetilde{\mathcal{Q}}^-(u_{lk})\mathcal{Q}^{++}_l(u_{lk})Q_\theta ^{[-q+1]}(u_{lk})}{\widetilde{\mathcal{Q}}^+(u_{lk})\mathcal{Q}^{--}_l(u_{lk})Q_\theta ^{[q-1]}(u_{lk})} \, ,
\nonumber \\
 \frac{\mathcal{Q}^-(u_{rk})}{\mathcal{Q}^+(u_{rk})}
 &=&-\frac{\widetilde{\mathcal{Q}}^+(u_{rk})\mathcal{Q}^{--}_r(u_{rk})Q_\theta ^{[-q-1]}(u_{rk})}{\widetilde{\mathcal{Q}}^-(u_{rk})\mathcal{Q}^{++}_r(u_{rk})Q_\theta ^{[q+1]}(u_{rk})}\,.
\end{eqnarray}
The additional contribution does not jeopardize the group-theory structure of the Bethe equations in the following five cases:
\begin{eqnarray}
 &&\begin{bmatrix}
 q_v  \\ 
  0 \\ 
  q_w \\ 
 \end{bmatrix}\longrightarrow 
  \begin{bmatrix}
 q_v  \\ 
  0 \\ 
  q_w \\ 
 \end{bmatrix},
  \qquad 
  \begin{bmatrix}
 0 \\ 
  q \\ 
  0 \\ 
 \end{bmatrix}\longrightarrow 
  \begin{bmatrix}
 q-1  \\ 
  -q \\ 
  q+1 \\ 
 \end{bmatrix},
   \qquad 
  \begin{bmatrix}
 1-q \\ 
  q \\ 
  0 \\ 
 \end{bmatrix}\longrightarrow 
  \begin{bmatrix}
 0  \\ 
  -q \\ 
  q+1 \\ 
 \end{bmatrix},
\nonumber \\
  &&\begin{bmatrix}
 0 \\ 
  q \\ 
  -1-q \\ 
 \end{bmatrix}\longrightarrow 
  \begin{bmatrix}
 q-1  \\ 
  -q \\ 
  0 \\ 
 \end{bmatrix},
   \qquad 
  \begin{bmatrix}
 1-q \\ 
  q \\ 
  -1-q \\ 
 \end{bmatrix}\longrightarrow 
  \begin{bmatrix}
 0  \\ 
  -q \\ 
  0 \\ 
 \end{bmatrix}.
\end{eqnarray}

\begin{figure}[t]
\begin{center}
 \centerline{\includegraphics[width=5cm]{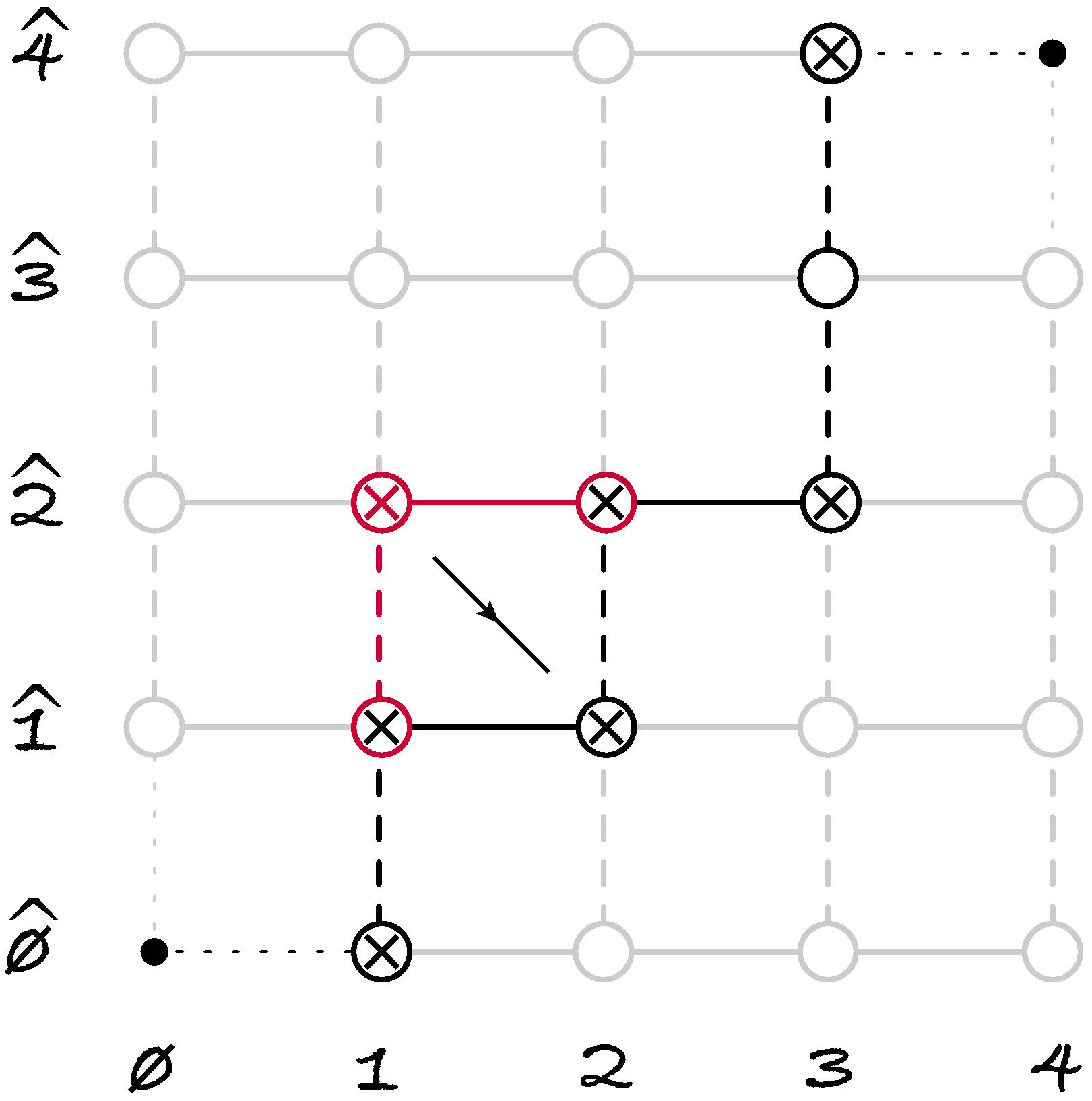}}
\caption{\label{Hasse}\small The Bethe equations are defined on a path connecting the two trivial Q-functions $\mathcal{Q}_{\emptyset|\widehat{\emptyset}}=1=\mathcal{Q}_{4|\widehat{4}}$. The horizontal direction roughly corresponds to $S^5$ and the vertical direction to $AdS_5$ in the dual string picture. The nodes at which the Dynkin diagram makes a turn are fermionic, and those which it passes straight are bosonic. The duality acts by flipping the path across a plaquette.}
\end{center}
\end{figure}

The fermionic duality reflects the non-uniqueness of the Cartan basis in a superalgebra \cite{Frappat:1996pb} and relates Bethe equations for Cartan matrices of different grading \cite{Tsuboi:1998ne}. It has multiple uses in the spin-chain description of the SYM spectrum \cite{Beisert:2005di,Kazakov:2007fy,Gromov:2014caa,Gromov:2017blm}. The Q-functions generated by the fermionic duality are labelled by two integers from $\emptyset$ to $4$: $\mathcal{Q}_{a|\widehat{a}}$, and can be placed at the vertices of a $5\times 5$ rectangle (fig.~\ref{Hasse})   \cite{Kazakov:2007fy}.  A particular grading is a path connecting the opposite corners. The nodes where the path turns are fermionic. The duality acts diagonally on each plaquette relating the Q-functions at the opposing corners.

This is not the end of the story, and it is important to mention that
the full set of Q-functions is defined on the Hasse diagram, with the duality equations acting along each square~\cite{Tsuboi:2009ud}.  This algebraic structure underlies the solution of the AdS/CFT spectral problem via the Quantum Spectral Curve \cite{Gromov:2014caa} and has been extensively studied from different 
angles~\cite{Gromov:2010km,Kazakov:2010iu,Bazhanov:2010jq,Frassek:2010ga,Kazakov:2015efa,Marboe:2016yyn,Gromov:2017blm,Kazakov:2018ugh}.
 While this extended structure is indispensable at the non-perturbative level, fermionic dualities alone are sufficient to solve the one-loop $\mathfrak{psu}(2,2|4)$ spin chain. For the Q-functions that fit on the square the fermionic duality relations are actually equivalent to the full set of Bethe equations \cite{Marboe:2016yyn}. 
 
At the level of the Q-functions no grading is distinguished. One can choose any path on the square or, better say, consider all the Q-functions on the same footing. The overlap formulae are formulated in a completely orthogonal way. They require fixing the grading and finding the Bethe roots. Ideally, we would like to have an invariant formulation where the path (the choice of grading) is not important at all. This we will not be able to achieve, but at least we will formulate the rules for how to transform overlaps from one grading to another, thus making the overlap formulae if not invariant under the fermionic duality then at least covariant.

\section{Determining Transformation Laws}
\label{sec:TransLaws}

In this section we determine how Gaudin superdeterminants \eqref{GaudinSuper} transform under fermionic duality transformations. The section consists of two parts. In the first part we focus on the dualization of non-momentum-carrying nodes. In the second part we then extend our results to momentum-carrying nodes.

\subsection{Non-Momentum-Carrying Nodes}

\subsubsection{Two-Node Example}
The simplest example of the fermionic duality occurs for the $\ocircle\!\!-\!\!\!-\!\otimes$ Dynkin diagram, where the bosonic node is momentum-carrying. The dual diagram is $\otimes\!\!-\!\!\!-\!\otimes$:
\begin{equation}
 M=\begin{bmatrix}
 2  & -1  \\ 
  -1 & 0 \\ 
 \end{bmatrix},\qquad \widetilde{M}=\begin{bmatrix}
 0  & 1  \\ 
  1 & 0 \\ 
 \end{bmatrix}.
\end{equation}
The Dynkin labels are the same in both cases:
\begin{equation}
 q=\widetilde{q}=\begin{bmatrix}
  1 \\ 
  0 \\ 
 \end{bmatrix}.
\end{equation}
The fermionic duality is expressed by the equation
\begin{equation}\label{fermionduality}
 Q_1^+-Q_1^-=iK_1u\widetilde{Q}_2Q_2 \, ,
\end{equation}
where $Q_a^{\pm}=Q_a(u \pm \frac{i}{2})$ are the reduced Baxter polynomials \eqref{eqn:ReducedBaxter} which only include paired roots. Indeed, the fermionic roots satisfy the equation
\begin{equation}
 1=\frac{Q_1^-(u_{2j})}{Q_1^+(u_{2j})} \,,
\end{equation}
and so do the roots of the polynomial $Q_1^+-Q_1^-$. The unused entries are the dual roots, which obviously satisfy the same Bethe equation. It is easy to see that trading $u_{2j}$ for $\widetilde{u}_{2j}$  cancels  self-scattering for $u_{1j}$ and flips the $u_1-u_2$ interaction, resulting the Cartan matrix $\widetilde{M}$.

We assume that the original roots are fully paired, i.e., $\left\{u_{aj},-u_{aj}\right\}$ for $j=1,..,K_a/2$ in order for the state to be bosonic. The dual roots then form $(K_1-K_2)/2-1$ pairs plus a zero root with the latter being the origin of the variable $u$ on the right-hand side of equation \eqref{fermionduality}. One advantage of using reduced Baxter polynomials is thus that zero roots are always clearly visible in the duality equations.

Note that although the original operator is bosonic, the dual operator is fermionic. For example, the state with two $u_1$ roots corresponds to 
\begin{equation}\label{OB}
 \mathcal{O}_B=\sum_{l=1}^{L-1}\mathop{\mathrm{tr}}XZ^{l-1}XZ^{L-l-1}\cos p\left(l-\frac{1}{2}\right) \, ,
\end{equation}
while the dual operator has the same $u_1$ roots but one additional $u_2$ root at zero. The corresponding operator reads
\begin{equation}\label{OF}
 \mathcal{O}_F=\sum_{l=1}^{L-1}\mathop{\mathrm{tr}}XZ^{l-1}\Psi Z^{L-l-1}\cos p\left(l-\frac{1}{2}\right) \, .
\end{equation}
The two operators are related  by supersymmetry and belong to the same multiplet, but the notion of highest weight changes with grading and while the first (bosonic) operator is primary for $M$, it becomes a descendant in the $\widetilde{M}$ grading. Consequently, overlaps with primaries will map to overlaps with descendants under the duality transformations.

Let us consider the simplest configuration $\left\{\left\{u_1,-u_1\right\},\left\{\right\}\right\}$ corresponding to the operator \eqref{OB} and its dual $\left\{\left\{u_1,-u_1\right\},\left\{0\right\}\right\}$. The Gaudin factors for this collection of roots are
\begin{equation}
 G^+=\frac{L}{u_1^2+\frac{1}{4}}\,,
 \qquad 
 G^-=\frac{L-1}{u_1^2+\frac{1}{4}}=\widetilde{G}^-\,,
 \qquad 
 \widetilde{G}^+=\frac{1}{u_1^2+\frac{1}{4}}\,\begin{bmatrix}
  L-1 & \sqrt{2} \\ 
  \sqrt{2} & -2 \\ 
 \end{bmatrix},
\end{equation}
and thus
\begin{equation}
\widetilde{\dd}=-\frac{2}{u_1^2+\frac{1}{4}}\,\dd \, .
\end{equation}
The general transformation rule reads
\begin{equation}
 \widetilde{\dd}=K_1\,\frac{\widetilde{Q}_2(0)Q_2(0)}{Q_1\left(\frac{i}{2}\right)}\,\dd \, ,.
\end{equation}
The above equation must be a general algebraic fact but was found numerically and holds semi-off-shell. The latter term means that the $u_1$ roots can be arbitrary numbers while the $u_2$ roots must be chosen such that the duality equation \eqref{fermionduality} is fulfilled. The Bethe equations for the $u_2$ roots follow as a consequence.

The overlap formulae will transform covariantly provided that $u_2$'s only enter through $Q_2(0)$ and $\dd$ and only in combination $Q_2(0)\dd$. The dual formula will then contain $\widetilde{\dd}/\widetilde{Q}_2(0)$. The factor $K_1$ takes into account that the original operator becomes a descendant in the dual frame.  Factors like that are indeed expected to appear in the overlaps of descendants (see appendix~A of \cite{deLeeuw:2017dkd}).

\subsubsection{Three-Node Example}

Next, we consider the $\mathfrak{su}(2|2)$ extension of the above Dynkin diagram which pictorially is given by 
$\ocircle\!\!-\!\!\!-\!\!\otimes\!\!-\!\!\!-\!\!\ocircle$,
where as before the left node is momentum-carrying. The dual diagram is 
$\otimes\!\!-\!\!\!-\!\!\otimes\!\!-\!\!\!-\!\!\otimes$. 
The corresponding Cartan matrices read:
\begin{equation}\label{O-X-O}
 M=\begin{bmatrix}
 2  & -1 & 0  \\ 
  -1 & 0 & 1 \\ 
    0 & 1 & -2 \\ 
 \end{bmatrix},\qquad \widetilde{M}=\begin{bmatrix}
 0  & 1 & 0  \\ 
  1 & 0 & -1 \\
  0 & -1 & 0 \\ 
 \end{bmatrix}.
\end{equation}
The Dynkin labels are the same in both cases:
\begin{equation}
 q=\widetilde{q}=\begin{bmatrix}
  1 \\ 
  0 \\ 
  0 \\
 \end{bmatrix}.
\end{equation}
Assuming that the roots at the neighboring nodes are fully paired, the fermionic duality is expressed by the equation
\begin{equation}\label{fermiondualitysu22}
 Q_1^-  Q_3^+ - Q_1^+  Q_3^-=i(K_3-K_1) u Q_2 \widetilde{Q}_2 \, ,
\end{equation}
where as before $Q_a^{\pm}=Q_a(u \pm \frac{i}{2})$ are the reduced Baxter polynomials \eqref{eqn:ReducedBaxter}. Trading the roots $u_{2j}$ for dual roots $\widetilde{u}_{2j}$ obviously cancels the self-scattering for $u_{1j}$ and $u_{3j}$ and flips the signs of the interactions resulting in the Cartan matrix $\widetilde{M}$.

Again, we assume that the original roots are fully paired, i.e., we also assume the $u_2$ roots to be paired. The dual roots then form $(K_1+K_3-K_2)/2-1$ pairs plus a zero root. In the considered case the general transformation rule for the Gaudin superdeterminant \eqref{GaudinSuper} is given by
\begin{equation}\label{exam}
 \widetilde{\dd}=(K_1-K_3)\,\frac{\widetilde{Q}_2(0)Q_2(0)}{Q_1\left(\frac{i}{2}\right)Q_3\left(\frac{i}{2}\right)}\,\dd \, ,
\end{equation}
which was found numerically and checked for various examples.

\subsubsection{General Case}

Let us finally turn to the most generic situation characterized by a fermionic, non-momentum-carrying node which has an arbitrary number of neighbors of arbitrary nature to both sides. Since the duality transformation only acts on nearest neighbors, we can account for this situation by considering the following parametric $3 \times 3$ Cartan matrix
\begin{align}
 M=\begin{bmatrix}
 \eta_2  & \eta_1 & 0 \\ 
  \eta_1 & 0 & -\eta_1\\  
   0 & -\eta_1 & \eta_3 \\
 \end{bmatrix} \, ,\qquad  q=\begin{bmatrix}
  V_l \\ 
  0 \\ 
  V_r \\
 \end{bmatrix} \, ,  \qquad 
 \begin{array}{l}
   \eta_1 \in \{-1,+1\} \\
   \eta_2 \in \{0,-2 \eta_1 \} \\
   \eta_3 \in \{0,2 \eta_1 \} 
    \end{array} \,,
\end{align}
which might be part of some bigger Cartan matrix. The nearest neighbors can either be bosonic or fermionic as parametrized by $\eta_2$ and $\eta_3$ and might or might not carry momentum depending on whether the Dynkin labels $V_l$ and $V_r$ are non-zero or not. The variable $\eta_1$ parametrizes the interactions between different families of Bethe roots. Dualizing the middle node maps the above Cartan matrix and Dynkin labels to
\begin{align}
\widetilde{M}=\begin{bmatrix}
 \eta_2+2 \eta_1 & -\eta_1 & 0 \\ 
  -\eta_1 & 0 & \eta_1\\  
   0 & \eta_1 & \eta_3-2 \eta_1 \\
 \end{bmatrix} \, ,\qquad  \widetilde{q}=q \, .
\end{align}
If all roots at the neighboring levels are fully paired, the fermionic duality is expressed by the equation
\begin{equation}\label{FDualityNMCGeneralPaired}
Q_l^- Q_r^+  -  Q_l^+ Q_r^-  =i \eta_1 ( K_r - K_l) u \,  Q_m \widetilde{Q}_m \, ,
\end{equation}
where $Q_{l,r}^{\pm}=Q_{l,r}(u \pm \eta_1 \frac{i}{2})$ are the reduced Baxter polynomials \eqref{eqn:ReducedBaxter} associated with the left and right node, respectively. Obviously, there is a zero root associated with the middle node in this case which can either be part of the original set of Bethe roots or the set of dual roots. When considering integrable overlaps one can also face a situation where there is a single unpaired zero root at one of the neighboring levels. In this case, the fermionic duality equation takes the form
\begin{equation}\label{FDualityNMCGeneralUnPaired}
\left(u \pm \eta_1 \tfrac{i}{2} \right) Q_l^- Q_r^+  -  \left(u \mp \eta_1 \tfrac{i}{2} \right) Q_l^+ Q_r^-  =i \eta_1 ( K_r - K_l)   Q_m \widetilde{Q}_m \, .
\end{equation}
Here, the upper signs pertain to the case of a zero root at the right node while the lower signs pertain to the case of a zero root at the left node.

Finally, we are now going to state how the Gaudin superdeterminant \eqref{GaudinSuper} transforms under the above transformation. We begin by focusing on the case where all roots are paired in the original grading. In this case, the dual roots form $(K_l+K_r-K_m)/2-1$ pairs plus a zero root and the transformation law reads
\begin{align}
\widetilde{\dd}=\mathrm{J}\,\dd \, ,
\label{eqn:TLawNMCgenericPaired}
\end{align}
where
\begin{align}
\mathrm{J}=(-\eta_1)^{K_l} \eta_1^{K_r}\, ( \eta_1 K_r - \eta_1 K_l) \, \frac{ Q_m(0) \widetilde{Q}_m(0)}{Q_l\left(\frac{i}{2}\right) Q_r\left(\frac{i}{2}\right)} \, .
\label{eqn:JacNMCGeneral}
\end{align}
As before, this result was found numerically and checked for various examples. If the zero root associated with the middle node is part of the original set of roots instead of the dual set, $\mathrm{J}$ must be replaced by $(-\mathrm{J})^{-1}$ which is consistent with the fact that the duality transformation needs to square to the identity.\footnote{Recall that the duality transformation flips the sign of $\eta_1$. The minus sign is hence necessary to ensure proper cancellation of pre-factors after the duality is applied twice.} Finally, we note that the transformation law \eqref{eqn:TLawNMCgenericPaired} also pertains to the situation where there is an unpaired zero root among the right or left set of Bethe roots. However, in this case the fermionic duality equation \eqref{FDualityNMCGeneralUnPaired} evaluated at $u=0$ in fact ensures that $\mathrm{J}=1$ so that effectively
\begin{align}
\widetilde{\dd}=\dd \, .
\label{eqn:TLawNMCgenericUnPaired}
\end{align}

\subsection{Momentum-Carrying Nodes}

\subsubsection{Two-Node Example}
Dualizing momentum-carrying nodes can potentially lead to different results. To address this issue we consider the Dynkin diagram $\otimes\!\!-\!\!\!-\!\ocircle$ and dualize the left node which is momentum-carrying in our case. The dual diagram is obviously $\otimes\!\!-\!\!\!-\!\otimes$. The corresponding Cartan matrices read
\begin{equation}
 M=\begin{bmatrix}
 0  & 1   \\ 
  1 & -2 \\  
 \end{bmatrix},\qquad \widetilde{M}=\begin{bmatrix}
 0  & -1   \\ 
  -1 & 0  \\
 \end{bmatrix}.
\end{equation}
The Dynkin labels are given by
\begin{equation}
 q=\begin{bmatrix}
  1 \\ 
  0 \\ 
 \end{bmatrix} \, , \qquad  \widetilde{q}=\begin{bmatrix}
  -1 \\ 
  0 \\ 
 \end{bmatrix} \, .
\end{equation}
The transformation law of Gaudin superdeterminants in general depends on the considered root configuration. In the original $\otimes\!\!\!-\!\!\!-\!\ocircle$ grading the momentum-carrying roots always have to come in pairs, while the roots at the auxiliary level can either all be paired or be paired up to a single unpaired zero root. In what follows we will treat these two cases separately.
\paragraph{Fully Paired Roots.}
In this paragraph we assume that the original roots are fully paired, i.e. $\left\{u_{aj},-u_{aj}\right\}$ for $j=1,..,K_a/2$. In this case, the fermionic duality is expressed by the equation
\begin{equation}\label{fermiondualityXOpaired}
 \left(u-\tfrac{i}{2}\right)^L  Q_2^+ -  \left(u+\tfrac{i}{2}\right)^L  Q_2^-=i(K_2-L) u \,  Q_1 \widetilde{Q}_1 \, .
\end{equation}
For even lengths the dual roots form $(L+K_2-K_1)/2-1$ pairs plus a zero root. In this case the general transformation rule for the Gaudin superdeterminant \eqref{GaudinSuper} is given by
\begin{equation}
\widetilde{\dd}=\left(2i\right)^{L}  (L-K_2)\,\frac{ Q_1(0) \widetilde{Q}_1(0)}{Q_2\left(\frac{i}{2}\right)}\,\dd \, ,
\label{eqn:translawXOOX}
\end{equation}
which was just as before found numerically. The combination of Baxter polynomials is obviously the same as in the non-momentum-carrying case.

\paragraph{Zero Root at the Auxiliary Level.}

The second situation to consider is characterized by a set of fully paired momentum-carrying roots while the auxiliary roots are only paired up to a single zero root, i.e.
\begin{align}
\left\{\left\{u_{j},-u_{j}\right\}_{j=1}^{\frac{K_1}{2}},\left\{y_{k},-y_{k},0\right\}_{k=1}^{\frac{K_2-1}{2}}\right\} \, ,
\end{align}
where $u_j$ are the momentum-carrying roots and $y_k$ are the auxiliary roots.
The fermionic duality equation reads
\begin{equation}\label{fermiondualityXOunpaired}
 \left(u-\tfrac{i}{2}\right)^L \left(u+\tfrac{i}{2}\right) Q_2^+ -  \left(u+\tfrac{i}{2}\right)^L  \left(u-\tfrac{i}{2}\right) Q_2^-=i (K_2-L) Q_1 \widetilde{Q}_1 \, .
\end{equation}
Note that the fermionic duality equation is in principle universal in the sense that it does not care about the different classes of root configurations. However, since we chose to work with reduced Baxter polynomials \eqref{eqn:ReducedBaxter}, the two equations \eqref{fermiondualityXOpaired} and \eqref{fermiondualityXOunpaired} look slightly different. Our notation, however, makes it clear that the situation where there is a zero root among the auxiliary roots needs special care as this zero inevitably leads to dual roots located at $\pm i/2$. Schematically, the dual configuration in the $\otimes\!\!-\!\!\!-\!\otimes$ grading thus looks as follows:
\begin{align}
\left\{\left\{\tfrac{i}{2},-\tfrac{i}{2},\tilde{u}_{j},-\tilde{u}_{j}\right\}_{j=2}^{\frac{\tilde{K}_1}{2}},\left\{ y_{k},-y_{k},0\right\}_{k=1}^{\frac{K_2-1}{2}}\right\} \, .
\end{align}
Naively, roots at $\pm i/2$ render the BAE as well as the overlap formulae divergent. For this reason, we first need to introduce an adequate regularization scheme. To establish such a scheme, we fist look at the BAE in the dual $\otimes\!\!-\!\!\!-\!\otimes$ grading:
\begin{align}
\label{eqn:BAEXX}
1&= \left(\frac{\tilde{u}_k+\frac{i}{2}}{\tilde{u}_k-\frac{i}{2}}\right)^{L} \,
\prod_{l=1}^{K_2} \frac{\tilde{u}_k-y_l-\frac{i}{2}}{\tilde{u}_k-y_l+\frac{i}{2}} \, , \nonumber \\
1&=\prod_{l=1}^{\tilde{K}_1} \,\,
\frac{y_k-\tilde{u}_l-\frac{i}{2}}{y_k-\tilde{u}_l+\frac{i}{2}} \, .
\end{align}
In order to regularize the singular roots, we make the ansatz \cite{Beisert:2003xu,Nepomechie:2013mua}
\begin{align}
\tilde{u}_1&=\tfrac{i}{2}+\varepsilon+ c_1 \varepsilon^L \, , \nonumber \\
\tilde{u}_2&=-\tfrac{i}{2}+\varepsilon+ c_2 \varepsilon^L \, , \nonumber \\
y_N&= \varepsilon \, ,
\label{eqn:XXregscheme}
\end{align}
where $N=\tilde{K_1}+K_2$ is the total number of roots and $c_1$ and $c_2$ are yet to be determined constants. Plugging the above expressions into the BAE \eqref{eqn:BAEXX} and requiring that these hold true in the limit $\varepsilon\rightarrow 0$ yields the following expressions for the two constants:
\begin{align}
c_1=(-i)^{L-1} \prod_{l \, \, \text{regular}} \frac{y_l-i}{y_l} \, , \qquad c_2=i^{L-1} \prod_{l \, \, \text{regular}} \frac{y_l+i}{y_l} \, .
\end{align}
In addition to the BAE, the Gaudin determinant needs to be regularized as well. However, we note that the regularization prescription \eqref{eqn:XXregscheme} is not symmetric and the regularized Gaudin determinant hence no longer factorizes. We will now describe how to overcome this issue. We begin by considering the full Gaudin determinant and set the singular roots as well as the zero root to their regularized values. The remaining roots stay undetermined for the moment. Expanding the matrix elements in $\varepsilon$ yields the following expression:
\begin{scriptsize}
\begin{align*}
\det(G)=\left|
\begin{array}{ccccccc}
-\frac{i}{c_1 \varepsilon^L}+\frac{L \, i}{\varepsilon} +\Phi _{1,1} & \Phi _{1,2} & \Phi _{1,3} & \ldots & \Phi _{1,N-1} & \frac{i}{c_1 \varepsilon^L} +\Phi _{1,N} \\
 \Phi _{2,1} & \frac{i}{c_2 \varepsilon^L}-\frac{L \, i}{\varepsilon} + \Phi _{2,2}& \Phi _{2,3} & \ldots  & \Phi _{2,N-1} & -\frac{i}{c_2 \varepsilon^L} +\Phi _{2,N} \\
 \Phi _{3,1} & \Phi _{3,2} & \Phi _{3,3} & \ldots  & \Phi _{3,N-1} & \Phi _{3,N} \\
 \vdots & \vdots & \vdots & \ddots & \vdots & \vdots \\
 \Phi _{N-1,1} & \Phi _{N-1,2} & \Phi _{N-1,3} & \ldots & \Phi _{N-1,N-1} & \Phi _{N-1,N} \\
 \frac{i}{c_1 \varepsilon^L} + \Phi _{N,1} & -\frac{i}{c_2 \varepsilon^L} + \Phi _{N,2} & \Phi _{N,3} & \ldots & \Phi _{N,N-1} & \frac{i (c_1-c_2)}{c_1 c_2 \varepsilon^L} +\Phi _{N,N} \\
\end{array}
\right| \, .
\end{align*}
\end{scriptsize}
Note that we only listed the divergent contributions explicitly while packaging all finite contributions into the expressions $\Phi_{i,j}$. In  parts, the latter are polynomials in $\varepsilon$ but for our purposes it is sufficient to consider them at $\varepsilon=0$. By definition, the $\Phi_{i,j}$ elements are thus independent of $\varepsilon$. We now perform row and column manipulations in such a way that all off-diagonal $\varepsilon^L$ poles are canceled. To achieve this, we first add the first and the second column to the last column. Then we add the first and the second row to the last row. After that, the leading divergences sit on the diagonal in the upper left corner and the Gaudin determinant takes the following form:
\begin{scriptsize}
\begin{align*}
\det(G)=\left|
\begin{array}{cc}
\begin{matrix}
-\frac{i}{c_1 \varepsilon^L}+\frac{L \, i}{\varepsilon} +\Phi _{1,1} \hspace{23pt} &  \Phi _{1,2} \hspace{10pt} \\
\Phi _{2,1} \hspace{23pt} & \frac{i}{c_2 \varepsilon^L}-\frac{L \, i}{\varepsilon} + \Phi _{2,2} \hspace{10pt}
\end{matrix}
&
\begin{matrix}
\frac{L \, i}{\varepsilon} +\Phi _{1,1}+\Phi _{1,2}+\Phi _{1,N} \\
- \frac{L \, i}{\varepsilon} +\Phi _{2,1}+\Phi _{2,2}+\Phi _{2,N} 
\end{matrix} \\
\begin{matrix}
\vdots & \vdots \\
\frac{L \, i}{\varepsilon} +\Phi _{1,1}+\Phi _{2,1}+\Phi _{N,1} & - \frac{L \, i}{\varepsilon} +\Phi _{1,2}+\Phi _{2,2}+\Phi _{N,2}
\end{matrix}
&
G^{\text{red}}
\end{array}
\right| \, ,
\end{align*}
\end{scriptsize}
where $G^{\text{red}}$ is the reduced Gaudin matrix
\begin{scriptsize}
\begin{align*}
G^{\text{red}}=
\left(
\begin{array}{ccccc}
 \Phi _{3,3} & \ldots & \Phi _{3,N-1} & \Phi _{3,1}+\Phi _{3,2}+\Phi _{3,N} \\
 \vdots & \ddots & \vdots & \vdots \\
 \Phi _{N-1,3} & \ldots & \Phi _{N-1,N-1} & \Phi _{N-1,1}+\Phi _{N-1,2}+\Phi _{N-1,N} \\
 \Phi _{1,3}+\Phi _{2,3}+\Phi _{N,3} & \ldots & \Phi _{1,N-1}+\Phi _{2,N-1}+\Phi _{N,N-1} &  G^{\text{red}}_{N,N}\\
\end{array}
\right) \, ,
\end{align*}
\end{scriptsize}
where
\begin{align}
G^{\text{red}}_{N,N}=\Phi _{1,1}+\Phi _{1,2}+\Phi _{1,N}+\Phi _{2,1}+\Phi _{2,2}+\Phi _{2,N}+\Phi _{N,1}+\Phi _{N,2}+\Phi _{N,N} \, .
\end{align}
Laplace expanding the above Gaudin determinant shows that the coefficient of the leading divergence is given by the determinant of the reduced Gaudin matrix (modulo the product of $c_1$ and $c_2$), i.e.
\begin{align}
\det(G)=\frac{\det(G^{\text{red}})}{c_1 c_2 \varepsilon^{2L}} + \mathcal{O}(\varepsilon^{-2L+2}) \, .
\end{align}
The key insight to make everything work out nicely is to treat the reduced Gaudin determinant as if it were the Gaudin determinant itself. If all the spectator roots (with respect to regularization) are fully paired, the reduced Gaudin determinant indeed factorizes, i.e.
\begin{align}
\det G^{\text{red}} = \det G^{\text{red}}_{+} \det G^{\text{red}}_{-} \, ,
\end{align}
leading to the following Gaudin superdeterminant:
\begin{align}
\widetilde{\dd}=\frac{\det G^{\text{red}}_{+}}{\det G^{\text{red}}_{-}}\, .
\end{align}
Having defined the regularized Gaudin superdeterminant in the above way, it is now a straightforward exercise to check that
\begin{align}
\widetilde{\dd}=\dd \, ,
\end{align}
which exactly mirrors the situation encountered for non-momentum-carrying roots \eqref{eqn:TLawNMCgenericUnPaired}.

\subsubsection{General Case}

We close this section by considering a more generic situation where the momentum-carrying node is embedded into a longer Dynkin diagram so that it has neighbors to both sides which are either bosonic or fermionic in nature. More precisely, we consider the BAE associated with the following parametric Cartan matrix
\begin{align}
 M=\begin{bmatrix}
 \eta_2  & \eta_1 & 0 \\ 
  \eta_1 & 0 & -\eta_1\\  
   0 & -\eta_1 & \eta_3 \\
 \end{bmatrix} \, ,\qquad  q=\begin{bmatrix}
  0 \\ 
  V \\ 
  0 \\
 \end{bmatrix} \, ,  \qquad 
 \begin{array}{l}
   \eta_1 \in \{-1,+1\} \\
   \eta_2 \in \{0,-2 \eta_1 \} \\
   \eta_3 \in \{0,2 \eta_1 \} 
    \end{array} \,,
\end{align}
with a fermionic middle node that carries momentum $(V \neq 0)$. Here, $\eta_2$ and $\eta_3$ parametrize neighboring nodes that can either be bosonic or fermionic, while $\eta_1$ parametrizes the interactions between different families of Bethe roots. One might think of this Cartan matrix as being part of a bigger Cartan matrix. Under a duality transformation of the momentum-carrying middle node the above Cartan matrix and Dynkin labels are mapped to
\begin{align}
\widetilde{M}=\begin{bmatrix}
 \eta_2+2 \eta_1 & -\eta_1 & 0 \\ 
  -\eta_1 & 0 & \eta_1\\  
   0 & \eta_1 & \eta_3-2 \eta_1 \\
 \end{bmatrix} \, ,\qquad  \widetilde{q}=\begin{bmatrix}
  V-\eta_1 \\ 
  -V \\ 
  V+\eta_1 \\
 \end{bmatrix} \, .
\end{align}
The fermionic duality is expressed by the equation
\begin{equation}\label{fermiondualityO(X)X(X)O}
P(u)=\left(u+ V \tfrac{i}{2}\right)^L  Q_l^- Q_r^+  -  \left(u- V \tfrac{i}{2}\right)^L  Q_l^+ Q_r^-  =i(V L -\eta_1 K_l+\eta_1 K_r) u \,  Q_m \widetilde{Q}_m ,
\end{equation}
where $Q_{l,r}^{\pm}=Q_{l,r}(u \pm \eta_1 \frac{i}{2})$ are the reduced Baxter polynomials associated with the left and right node, respectively. The original roots are assumed to be fully paired, i.e., $\left\{u_{aj},-u_{aj}\right\}$ for $j=1,..,K_a/2$ so that for even lengths the dual roots form $(L+K_l+K_r-K_m)/2-1$ pairs plus a zero root. 

An important point to note concerns the total momentum phase of a Bethe state in the above set-up. Since the dual roots inevitably contain an uncompensated zero root, one may wonder about the fate of the zero-momentum condition which apparently seems to evaluate to $-1$ after the dualization. However, from \eqref{fermiondualityO(X)X(X)O} it follows that
\begin{align}
\frac{P(+V \tfrac{i}{2})}{P(-V \tfrac{i}{2})}&=(-1)^{-L-1} \prod\limits_{i=1}^{K_l} \frac{u_{l,i}-(V-\eta_1) \tfrac{i}{2} }{u_{l,i}+ (V-\eta_1) \tfrac{i}{2}} \prod\limits_{j=1}^{K_r} \frac{u_{r,j}-(V+\eta_1) \tfrac{i}{2} }{u_{r,j}+ (V+\eta_1) \tfrac{i}{2}} \nonumber \\
&=\prod\limits_{k=1}^{K_m} \frac{u_{m,k}-V \tfrac{i}{2} }{u_{m,k}+ V \tfrac{i}{2}} \prod\limits_{k'=1}^{\tilde{K}_m} \frac{\tilde{u}_{m,k'}-V \tfrac{i}{2} }{\tilde{u}_{m,k'}+ V \tfrac{i}{2}}
\end{align}
from which we conclude that
\begin{align}
\exp(i \mathrm{P})&=\prod\limits_{k=1}^{K_m} \frac{u_{m,k}+V \tfrac{i}{2} }{u_{m,k}- V \tfrac{i}{2}}  \\
&=(-1)^{L+1} \prod\limits_{i=1}^{K_l} \frac{u_{l,i}+(V-\eta_1) \tfrac{i}{2} }{u_{l,i}- (V-\eta_1)  \tfrac{i}{2}} \prod\limits_{k'=1}^{\tilde{K}_m} \frac{\tilde{u}_{m,k'}-V \tfrac{i}{2} }{\tilde{u}_{m,k'}+ V \tfrac{i}{2}} \prod\limits_{j=1}^{K_r} \frac{u_{r,j}+(V+\eta_1) \tfrac{i}{2} }{u_{r,j}- (V+\eta_1) \tfrac{i}{2}} \nonumber \, .
\end{align}
Since the duality transformation interchanges the role of the vacuum and its excitations, the exchange statistics of the spin vacuum is flipped after the transformation. This is reflected by the sign $(-1)^{L+1}$ which nicely compensates the $-1$ stemming from the zero root, see \cite{Beisert:2005di} for more details.

In the considered case the general transformation rule for the Gaudin superdeterminants \eqref{GaudinSuper} is given by
\begin{equation}
\widetilde{\dd}=\left(\frac{2 i}{V}\right)^{L}\, (V L -\eta_1 K_l+\eta_1 K_r) \, \frac{ Q_m(0) \widetilde{Q}_m(0)}{Q_l\left(\frac{i}{2}\right) Q_r\left(\frac{i}{2}\right)}\,\dd \, ,
\label{eqn:translawMCgeneric}
\end{equation}
which was again found numerically. The combination of reduced Baxter polynomials is thus the same as in the non-momentum-carrying case. Since the transformation formulae in the momentum-carrying and the non-momentum-carrying case are very similar, we strongly suspect the equation 
\begin{align}
\widetilde{\dd}=\dd \, ,
\end{align}
to hold for root configurations containing a neighboring unpaired zero root once proper regularization has been performed. However, we have only checked the last equation for the above example.

\section{Dualizing Overlap Formulae \label{sec:DualizingOF}}

In this section we apply the above insights to study how overlap formulae transform under a change of grading of the underlying algebra. We begin by working through an $\mathfrak{su}(2|2)$ example in great detail to illustrate the procedure. Finally, we leverage the newly gained insights to give a concise graphical procedure for how to transform overlap formulae between any two gradings.  

\subsection{Dualizing $\mathfrak{su}(2|2)$ Overlaps}
\label{translation}
We begin by focusing on the $\mathfrak{su}(2|2)$ case and consider the 
 $\otimes\!\!-\!\!\!-\!\!\ocircle\!\!-\!\!\!-\!\!\otimes$
grading as well as the 
$\ocircle\!\!-\!\!\!-\!\!\otimes\!\!-\!\!\!-\!\!\ocircle$
grading, where the left node is momentum-carrying in both cases. Both gradings are connected by fermionic duality transformations: going from 
 $\otimes\!\!-\!\!\!-\!\!\ocircle\!\!-\!\!\!-\!\!\otimes$
to 
$\ocircle\!\!-\!\!\!-\!\!\otimes\!\!-\!\!\!-\!\!\ocircle$
is achieved by first dualizing the third node and then dualizing the second node.

In the  $\otimes\!\!-\!\!\!-\!\!\ocircle\!\!-\!\!\!-\!\!\otimes$ grading the overlap formula for valence bond states of the form 
\begin{align}
 \left\langle B\right|=\bigl(\left\langle ZZ \right|+\left\langle XX \right| +\left\langle \uparrow \downarrow \right| - \left\langle\downarrow\uparrow  \right| \bigr) ^{\otimes \frac{L}{2}} \, , \label{eqn:FBoundary}
\end{align}
reads
\begin{equation}
\label{eqn:generalformulasu22}
\frac{ \left\langle B  | {\bf u_1}, {\bf u_2}, {\bf u_3}  \right\rangle}{\left\langle  {\bf u_1}, {\bf u_2}, {\bf u_3} |   {\bf u_1}, {\bf u_2}, {\bf u_3} \right\rangle^{1/2}}=
\sqrt{\frac{Q_{1}(0) Q_{3}(0)} {Q_{2}(0) Q_{2}\left(\frac{i}{2}\right) }\,\frac{\det G_+}{\det G_-}}\, .
\end{equation}
This formula slightly extends a result reported in \cite{Kristjansen:2020mhn}. Here, $u_1, u_2, u_3$ are the three families of roots and $Q_{i}$ are the associated reduced Baxter polynomials, i.e., zero roots simply contribute a factor of $1$. There are two root configurations that need to be considered: 
\begin{align}
&(1.1) \text{ the number of roots } K_1, K_2 \text{ and } K_3 \text{ are all even}  \nonumber\\
&(1.2)\, K_1 \text{ and } K_3 \text{ are even while } K_2 \text{ is odd (zero root at the second level)} \nonumber
\end{align}

We begin by dualizing the third node of the Dynkin diagram
 $\otimes\!\!-\!\!\!-\!\!\ocircle\!\!-\!\!\!-\!\!\otimes$
which results in 
  $\otimes\!\!-\!\!\!-\!\!\otimes\!\!-\!\!\!-\!\!\otimes$. The Cartan matrices read
\begin{equation}\label{X-O-X}
 M_{XOX}=\begin{bmatrix}
 0  & 1 & 0  \\ 
  1 & -2 & 1 \\ 
    0 & 1 & 0 \\ 
 \end{bmatrix},\qquad \widetilde{M}_{XXX}=\begin{bmatrix}
 0  & 1 & 0  \\ 
  1 & 0 & -1 \\
  0 & -1 & 0 \\ 
 \end{bmatrix}.
\end{equation}
The Dynkin labels are the same in both cases $q=\widetilde{q}=(1,0,0)$.  Since we are using reduced Baxter polynomials \eqref{eqn:ReducedBaxter}, the duality equation has to be phrased in a slightly different fashion depending on the considered root configuration:
\begin{align}
 &(1.1) \hspace{2cm} Q_2^+-Q_2^-=i K_2 u \widetilde{Q}_3 Q_3 \, , \nonumber \\
 &(1.2) \hspace{1cm}\left(u+\tfrac{i}{2} \right) Q_2^+ - \left(u-\tfrac{i}{2} \right)Q_2^-=i K_2  \widetilde{Q}_3 Q_3 \, .
\end{align}
The number of dual roots $\widetilde{u_3}$ is given by
\begin{align}
\widetilde{K_3}=K_2-K_3-1
\end{align}
and is therefore odd for an even number of roots $K_2$ $(1.1)$ and even for an odd number of roots $K_2$ $(1.2)$. For the transformation law of the Gaudin superdeterminants we find
\begin{align}
&(1.1) \; \widetilde{\dd}\left(\widetilde{M}_{XXX},u_1,u_2,\{\widetilde{u_3},0\}\right)\!=- K_2\,\frac{\widetilde{Q}_3(0)Q_3(0)}{Q_2\left(\frac{i}{2}\right)}\,\dd\left(M_{XOX},u_1,u_2,u_3\right)  , \nonumber\\
&(1.2) \; \widetilde{\dd}\left(\widetilde{M}_{XXX},u_1,\{u_2,0\},\widetilde{u_3} \right)\!=K_2\,\frac{\widetilde{Q}_3(0)Q_3(0)}{Q_2\left(\frac{i}{2}\right)}\,\dd\left(M_{XOX},u_1,\{u_2,0\},u_3\right)  ,
\label{eqn:DXOXtoXXX}
\end{align}
where we have listed the zero roots explicitly for maximal clarity. Obviously, both formulae only differ by a sign. The pre-factor in the lower equation is actually equal to $1$ as explained below equation \eqref{eqn:TLawNMCgenericPaired} but we prefer to keep it for later convenience.

The second step consists of dualizing the second node of the  $\otimes\!\!-\!\!\!-\!\!\otimes\!\!-\!\!\!-\!\!\otimes$ Dynkin diagram so that the final diagram reads  $\ocircle\!\!-\!\!\!-\!\!\otimes\!\!-\!\!\!-\!\!\ocircle$  corresponding to the Cartan matrix
\begin{equation}\label{CarMatOXO}
 \widetilde{\widetilde{M}}_{OXO}=\begin{bmatrix}
 2  & -1 & 0  \\ 
  -1 & 0 & 1 \\ 
    0 & 1 & -2 \\ 
 \end{bmatrix} \, ,
\end{equation}
while the Dynkin labels remain unchanged. Again, there are two different root configurations that we need to consider
\begin{align}
&(2.1) \; K_1, K_2 \text{ are even while } \widetilde{K}_3 \text{ is odd} \, ,  \nonumber\\
&(2.2)\;  K_1 \text{ and } \widetilde{K}_3 \text{ are even while } K_2 \text{ is odd} \, . \nonumber
\end{align}
The Bethe roots are dualized with the help of the equations
\begin{align}
\label{eqn:2nddualization}
&(2.1) \hspace{1cm}\left(u-\tfrac{i}{2} \right) Q_1^+  \widetilde{Q}_3^- - \left(u+\tfrac{i}{2}\right) Q_1^-  \widetilde{Q}_3^+=i(K_1-\widetilde{K}_3) Q_2 \widetilde{Q}_2 \, , \nonumber \\
 &(2.2) \hspace{2.2cm} Q_1^+  \widetilde{Q}_3^- -  Q_1^-  \widetilde{Q}_3^+= i(K_1-\widetilde{K}_3) u Q_2 \widetilde{Q}_2 \, .
\end{align}
The number of dual roots is given by
\begin{align}
\widetilde{K}_2=K_1+\widetilde{K}_3-K_2-1 \, .
\end{align}
For the transformation law of the Gaudin superdeterminants we find
\begin{align}
&(2.1) \; \widetilde{\widetilde{\dd}}\left(\widetilde{\widetilde{M}}_{OXO},u_1,\widetilde{u}_2,\{\widetilde{u_3},0\}\right)=- (K_1-\widetilde{K}_3)\,\frac{\widetilde{Q}_2(0)Q_2(0)}{Q_1\left(\frac{i}{2}\right) \widetilde{Q}_3\left(\frac{i}{2}\right)} \nonumber \\
&\hspace{6cm}\widetilde{\dd}\left(\widetilde{M}_{XXX},u_1,u_2,\{\widetilde{u_3},0\}\right) \, ,\nonumber \\
&(2.2) \; \widetilde{\widetilde{\dd}}\left(\widetilde{\widetilde{M}}_{OXO},u_1,\widetilde{u}_2,\widetilde{u_3}\right)=\frac{1}{(K_1-\widetilde{K}_3)}\,\frac{Q_1\left(\frac{i}{2}\right) \widetilde{Q}_3\left(\frac{i}{2}\right)}{\widetilde{Q}_2(0)Q_2(0)} \nonumber \\
&\hspace{5.2cm}\widetilde{\dd}\left(\widetilde{M}_{XXX},u_1,\{u_2,0\},\widetilde{u_3}\right) \, ,
\label{eqn:DXXXtoOXO}
\end{align}
where once again we have listed the zero roots explicitly for maximal clarity.

Combining the results \eqref{eqn:DXOXtoXXX} and \eqref{eqn:DXXXtoOXO} yields
\begin{align}
&\widetilde{\widetilde{\dd}}\left(\widetilde{\widetilde{M}}_{OXO},u_1,\widetilde{u}_2,\{\widetilde{u_3},0\}\right)=K_2\, (K_1-\widetilde{K}_3)\,\frac{\widetilde{Q}_2(0)\widetilde{Q}_3(0)}{\widetilde{Q}_3\left(\frac{i}{2}\right)} \frac{Q_2(0) Q_3(0)}{Q_1\left(\frac{i}{2}\right)  Q_2\left(\frac{i}{2}\right)} \nonumber \\
&\hspace{5cm}\dd\left(M_{XOX},u_1,u_2,u_3\right) \, ,\nonumber \\
&\widetilde{\widetilde{\dd}}\left(\widetilde{\widetilde{M}}_{OXO},u_1,\widetilde{u}_2,\widetilde{u_3}\right)=\frac{K_2}{(K_1-\widetilde{K}_3)}\,\frac{Q_1\left(\frac{i}{2}\right)}{Q_1(0)}\frac{\widetilde{Q}_3(0)\widetilde{Q}_3\left(\frac{i}{2}\right)}{\widetilde{Q}_2(0)}\frac{Q_1(0)Q_3(0)}{Q_2(0) Q_2\left(\frac{i}{2}\right)} \nonumber \\
&\hspace{4.2cm}\dd\left(M_{XOX},u_1,\{u_2,0\},u_3\right) \, .
\label{eqn:SGTrafoXOXtoOXO}
\end{align}
In order to see that both formulae actually agree, we evaluate the upper equation of \eqref{eqn:2nddualization} at $u=0$ and find
\begin{align}
 Q_1\left(\tfrac{i}{2}\right)  \widetilde{Q}_3\left(\tfrac{i}{2}\right) =-(K_1-\widetilde{K}_3) Q_2(0) \widetilde{Q}_2(0) \, ,
\end{align}
where we have used that $Q_i\left(-\tfrac{i}{2}\right)=Q_i\left(\tfrac{i}{2}\right)$. Using this relation the upper transformation law in equation \eqref{eqn:SGTrafoXOXtoOXO} can be rewritten as
\begin{align}
&\widetilde{\widetilde{\dd}}\left(\widetilde{\widetilde{M}}_{OXO},u_1,\widetilde{u}_2,\{\widetilde{u_3},0\}\right)=\frac{K_2}{(K_1-\widetilde{K}_3)}\,\frac{Q_1\left(\frac{i}{2}\right)}{Q_1(0)}\frac{\widetilde{Q}_3(0)\widetilde{Q}_3\left(\frac{i}{2}\right)}{\widetilde{Q}_2(0)}\frac{Q_1(0)Q_3(0)}{Q_2(0) Q_2\left(\frac{i}{2}\right)} \nonumber \\
&\hspace{5cm}\dd\left(M_{XOX},u_1,u_2,u_3\right) \, ,
\end{align}
proving that both transformation laws in equation \eqref{eqn:SGTrafoXOXtoOXO} are in fact equivalent. For both root configurations that are of interest to us we therefore obtain
\begin{align}
\frac{Q_1(0)Q_3(0)}{Q_2(0) Q_2\left(\frac{i}{2}\right)} \dd\left(M_{XOX},u_1,u_2,u_3\right)&=
\frac{K_1-\widetilde{K}_3}{K_1+\widetilde{K}_3-\widetilde{K}_2-1}\frac{Q_1(0)}{Q_1\left(\frac{i}{2}\right)}\frac{\widetilde{Q}_2(0)}{\widetilde{Q}_3(0)\widetilde{Q}_3\left(\frac{i}{2}\right)}\nonumber \\
&\hspace{0.5cm} \widetilde{\widetilde{\dd}}\left(\widetilde{\widetilde{M}}_{OXO},u_1,\widetilde{u}_2,\widetilde{u_3}\right) \, ,
\end{align}
where we have now dropped the notation highlighting the zero roots. This result is in perfect agreement with \eqref{D3D5-Alt} up to the factor which accounts for the fact that the state is a descendant in the new grading.

\subsection{Dualizing $\mathfrak{psu}(2,2|4)$ Overlaps \label{psu224}}

\begin{figure}[t]
\begin{center}
 \centerline{\includegraphics[width=8cm]{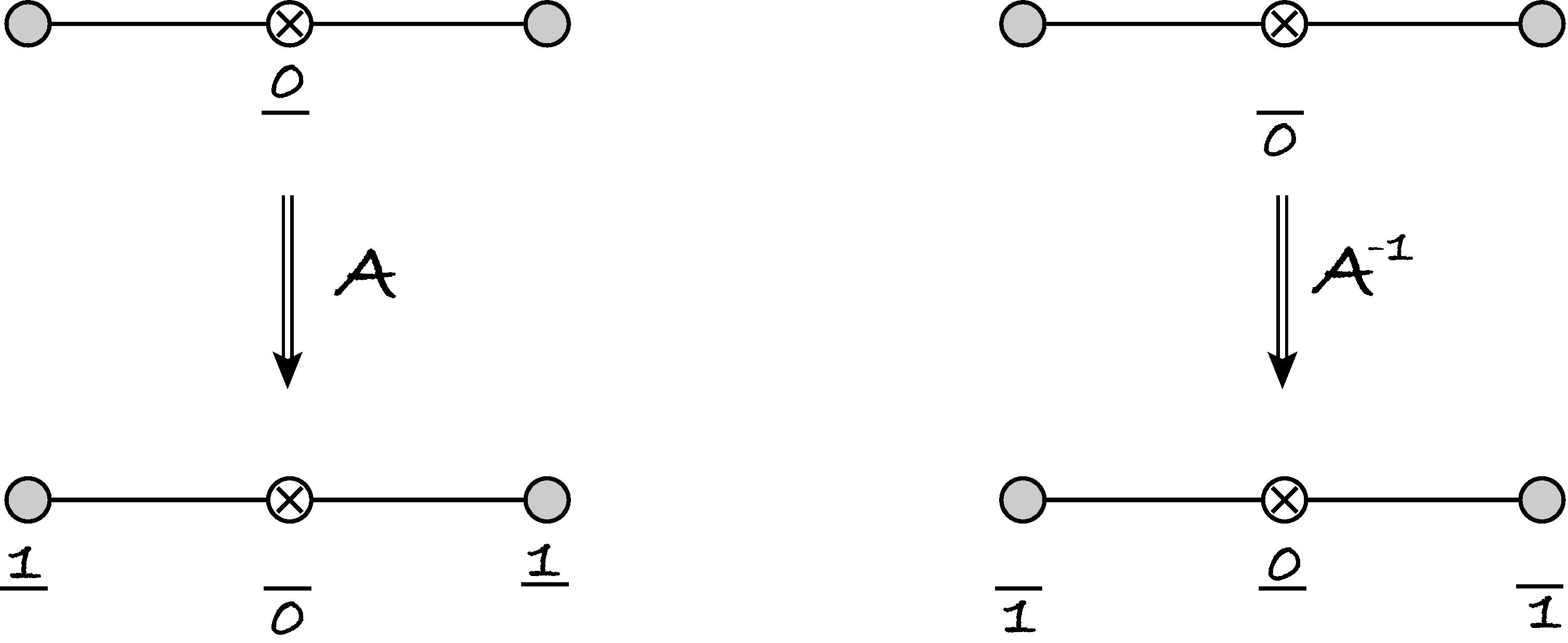}}
\caption{\label{Rules}\small Transformation rules for the overlaps.}
\end{center}
\end{figure}

Suppose duality on a fermionic node $a$ maps Dynkin diagram $M$ to $\widetilde{M}$, and an explicit expression for an overlap is known in the original grading. How will the overlap formula look in the grading $\widetilde{M}$? From the previous sections we know that the action of the duality on the Gaudin superdeterminant produces a Jacobian\footnote{This may look different from (\ref{exam}), but the roots only transform on the middle node, so $\widetilde{Q}_{a\pm 1}=Q_{a\pm 1}$.}
\begin{equation}
\dd\propto\frac{\widetilde{Q}_{a-1}\left(\frac{i}{2}\right)\widetilde{Q}_{a+1}\left(\frac{i}{2}\right)}{\widetilde{Q}_a(0)Q_a(0)}\,\widetilde{\dd}.
\end{equation}
Upon substituting this into the overlap formula, the Q-functions in the pre-factor
re-arrange themselves nicely if and only if 
the original pre-factor contained $Q_a(0)$, and no other  instances of $Q_a$. Then  $Q_a(0)$ flips to $1/\widetilde{Q}_a(0)$ and additional factors of $\widetilde{Q}_{a\pm 1}(i/2)$ appear on the adjacent nodes. Using graphic notations of fig.~\ref{gnot}, the resulting transformation rule is shown in fig.~\ref{Rules} on the left. 

When the pre-factor contains $1/Q_a(0)$ the overlap formula will also transform nicely. The original grading then corresponds to $\widetilde{M}$ and this results in the inverse transformation illustrated in  fig.~\ref{Rules} on the right. These are the two basic rules for transforming the overlaps. They are obviously quite restrictive. The only fermionic Q-functions allowed are $[Q_a(0)]^{\pm 1}$. The D3-D5 overlap (\ref{D3D5-Alt}) in fig.~\ref{Dynkin-Alt} is consistent with this requirement, but one should remember that duality changes bosonic nodes into fermionic and new restrictions arise.
It is quite remarkable that application of all possible duality transformations does not jeopardize the strict requirement imposed on the fermionic Q-functions.
We are going to demonstrate this shortly.

\begin{figure}[t]
\begin{center}
 \centerline{\includegraphics[height=5cm]{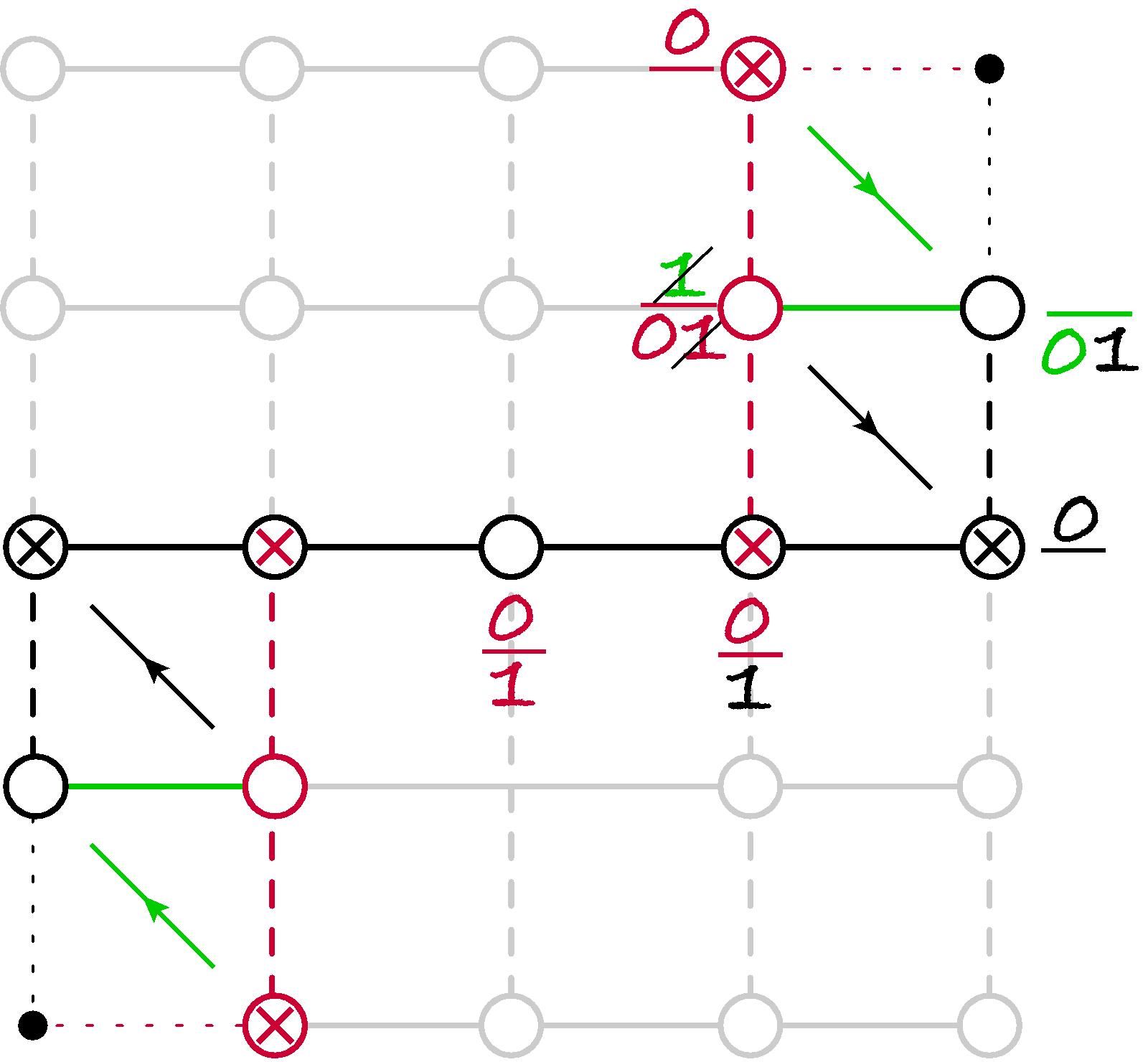}}
\caption{\label{Alt-to-SO6}\small Duality transformations from the 
alternating grading to the Beauty diagram: red to green to black.}
\end{center}
\end{figure}

\begin{figure}[t]
\begin{center} 
 \subfigure[]{
   \includegraphics[height=1.5cm] {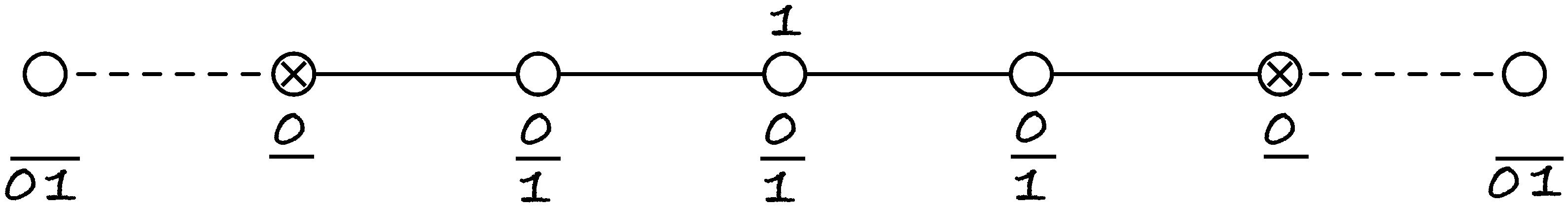}
   \label{Dynkin-SO6}
 }
 \subfigure[]{
   \includegraphics[height=1.5cm] {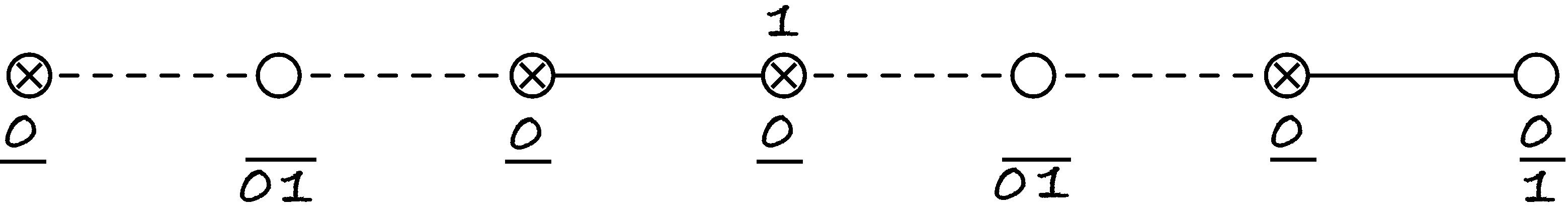}
   \label{Dynkin-F}
 }
 \subfigure[]{
   \includegraphics[height=1.5cm] {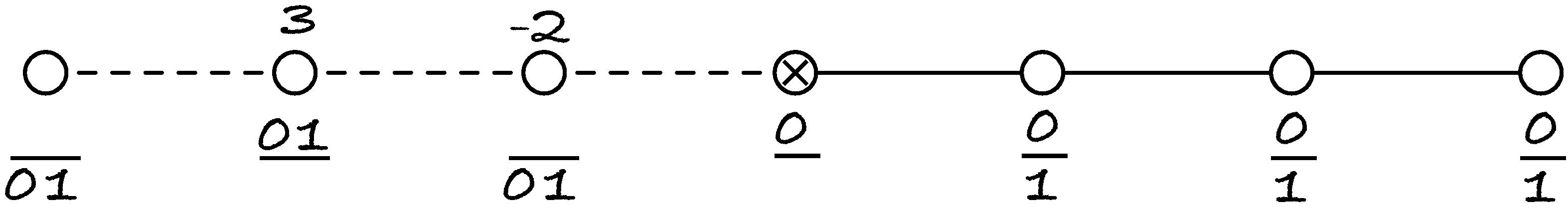}
   \label{Dynkin-Beast}
 }
\caption{\label{Dynkin-diagrams}\small 
Overlap formulae in different gradings, in graphic notations of fig.~\ref{gnot}: 
(a) for the Beauty diagram; 
(b) in the grading where a fermionic node is momentum-carrying;
(c) in the grading corresponding to the distinguished Dynkin diagram of $\mathfrak{psu}(2,2|4)$ (the Beast diagram).
}
\end{center}
\end{figure}

We are going to apply transformation rules to the overlap formula (\ref{D3D5-Alt}), written for the alternating Dynkin diagram in fig.~\ref{Dynkin-Alt}. The first example is given in~fig.~\ref{Alt-to-SO6}, which transforms the overlap to the $SO(6)$-friendly grading (dubbed in \cite{Beisert:2003yb} the "Beauty" diagram). The result is shown in fig.~\ref{Dynkin-SO6}. When restricted to the $SO(6)$ sector (the three middle nodes), the result agrees with the overlap derived in \cite{Kristjansen:2020mhn} thus establishing a link between the bootstrap approach of \cite{Gombor:2020kgu} and direct field-theory calculations.
 
\begin{figure}[t]
\begin{center}
 \centerline{\includegraphics[height=5.5cm]{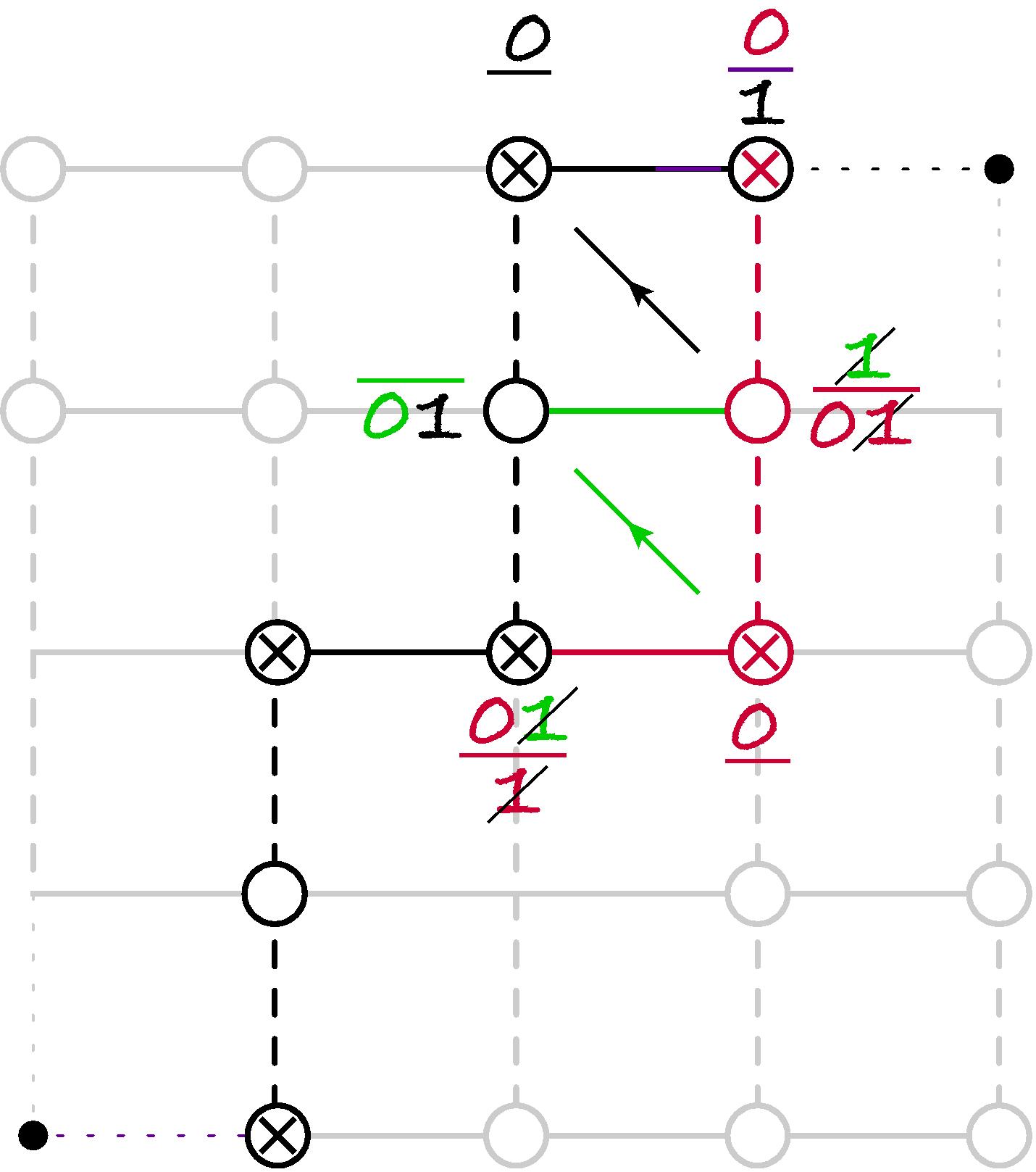}}
\caption{ \label{Alt-to-F}\small A graphic representation of the transformation  discussed in sec.~\ref{translation} from the alternating to the fermionic grading:  red to green to black.}
\end{center}
\end{figure}

Another case considered in \cite{Kristjansen:2020mhn} is an $SU(2|1)$ subsector of fermionic excitations $\Psi _\alpha $ on top of the vacuum composed of $Z$s. Comparison to the general formula \cite{Gombor:2020kgu} requires a chain of duality transformations illustrated in fig.~\ref{Alt-to-F}. This results in the overlap formula for the grading with a fermionic momentum-carrying node in fig.~\ref{Dynkin-F}. The $SU(2|1)$ subsector corresponds to the central node plus the one immediately to the right. The overlap again agrees with the results of the direct calculation \cite{Kristjansen:2020mhn}.

\begin{figure}[t]
\begin{center}
 \centerline{\includegraphics[height=6cm]{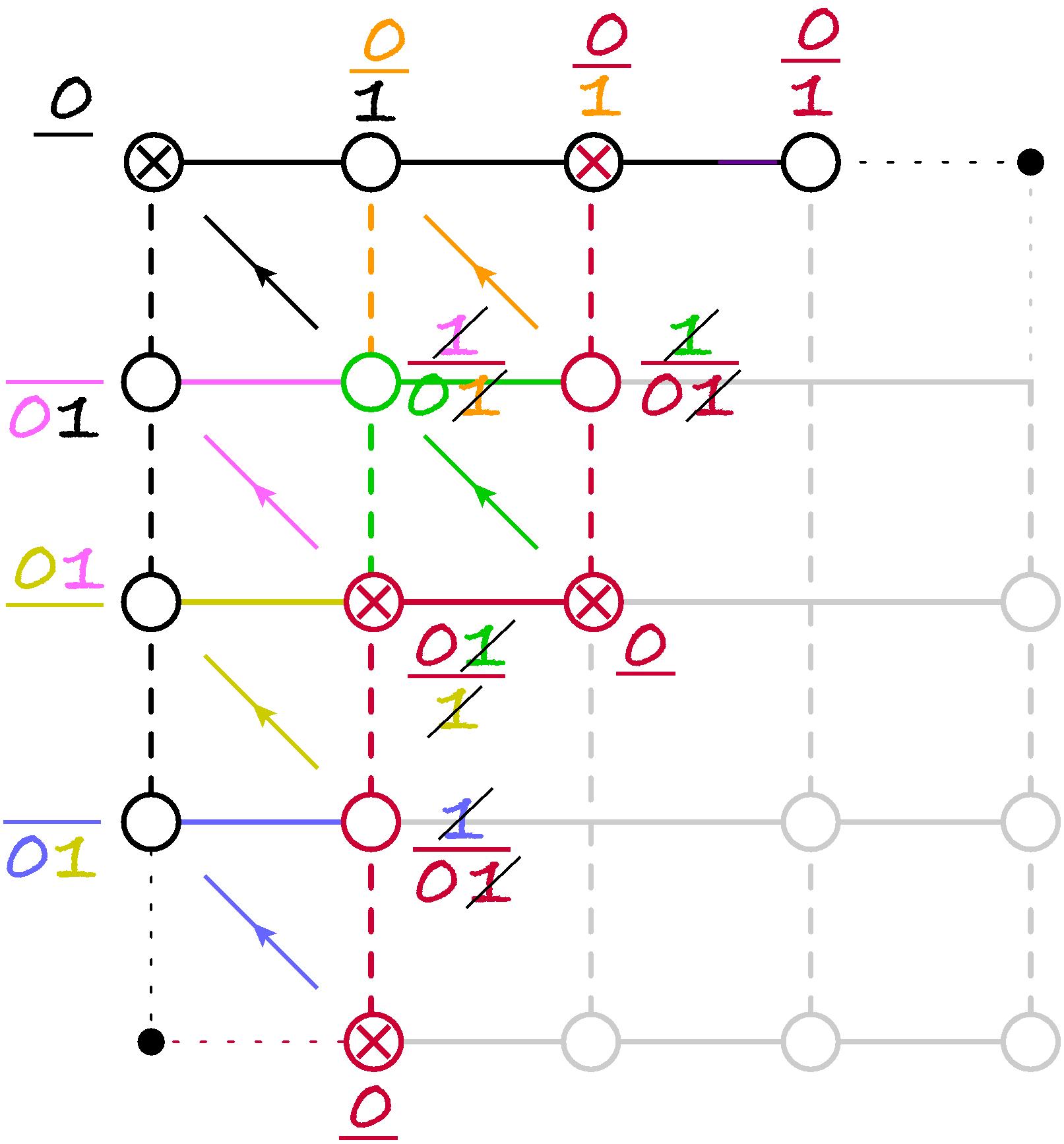}}
\caption{ \label{F-to-Beast}\small From fermion to Beast: red to green to orange to blue to jade to purple to black.}
\end{center}
\end{figure}

Finally we can transform the overlap to the distinguished Dynkin diagram of $\mathfrak{psu}(2,2|4)$, called  the "Beast" diagram in \cite{Beisert:2003yb}. The chain of dualities is shown in fig.~\ref{F-to-Beast} and results in the overlap illustrated in~\ref{Dynkin-Beast}. We can again make contact with \cite{Kristjansen:2020mhn} by considering the gluonic spin-1 $\mathfrak{su}(2)$ subsector formed by the self-dual components of the field strength. The Bethe ansatz for this spin chain is obtained by restricting to the momentum-carrying bosonic node in fig.~\ref{Dynkin-Beast}. The overlap with the D3-D5 state is proportional to $1/Q(0)Q(i/2)$, in agreement with the direct inspection of the spin chain, where the boundary state is the N\'eel-type $SU(2)$ singlet  \cite{Kristjansen:2020mhn}.

Apart from complete agreement with the direct field-theory computations, our findings reveal remarkable internal consistency of the overlap formula (\ref{D3D5-Alt}).  The fermionic duality happens to work as a miracle, enabling all elementary moves along the way. This is a very strong consistency condition, and we cannot exclude that duality covariance  fixes the overlap formula completely. 

\section{Conclusion\label{conclusion}}

Recent results on integrable one-point functions in domain wall versions of ${\cal N}=4$ SYM~\cite{Gombor:2020auk,Kristjansen:2020mhn}  have triggered a need to know how overlap formulae for super spin chains can be translated between different gradings of the underlying super Lie algebra. 
We recall that one-point functions in the D3-D5 defect set-up are expressed as overlaps between Bethe eigenstates and matrix product states for $k\geq 2$ and as overlaps between Bethe eigenstates and valence bond states for $k=1$. Evoking bootstrap arguments reference~\cite{Gombor:2020auk}  presented a closed formula for   the one-point functions of the D3-D5 set-up for $k\geq 2$ with the fields involved  characterized by quantum numbers referring to the alternating grading of the underlying $\mathfrak{psu}(2,2|4)$ algebra. Earlier an expression for the tree level one point functions of all scalar operators for $k\geq2 $ was found in the grading corresponding to the so-called beauty Dynkin diagram of the Lie algebra~\cite{deLeeuw:2018mkd}.  Furthermore, using the $k=1$ version of the D3-D5 set-up as a short cut for going beyond the scalar sector, one-point function formulae for certain fermionic as well as gluonic operators were derived in~\cite{Kristjansen:2020mhn} where it was also shown that for the simplest subset of scalar operators the $k=1$ result followed by analytical continuation from $k\geq2$.  In order to translate the expression of~\cite{Gombor:2020auk} to different gradings of the super Lie algebra and in particular to show its compatibility with the other results mentioned above it is imperative to know how the key component of all overlap formulae, the super determinant of the Gaudin matrix looks in different gradings. 
By means of consistency and covariance arguments, aided by numerical investigations, we have identified the transformation properties of the super determinant of the Gaudin matrix under fermionic dualities which allow one to pass between the different possible gradings of superalgebras of the $\mathfrak{gl}(M|N)$ type. Applied to the case of $\mathfrak{psu}(2,2|4)$ our transformation rules have allowed us to demonstrate the compatibility of results of~\cite{Gombor:2020auk} with all earlier results and to express the one-point functions 
in any grading of the super Lie algebra.

The fermionic dualities considered here constitute only a sub-set of duality transformations possible for the Bethe equations corresponding to an integrable spin chain
based on a super Lie algebra.  In general, the possible sets of Bethe equations and QQ-relations are encoded in the Hasse diagram which for the $\mathfrak{psu}(2,2|4)$
spin chain involves 256 nodes~\cite{Tsuboi:2009ud}. In addition to the fermionic dualities  the Hasse diagram entails two series of bosonic dualities. It would be very interesting if the
transformation properties of the super determinant of the Gaudin matrix, discovered here, could be generalized to the full Hasse diagram. We plan to address this question in future
work~\cite{inprogress}.
Likewise, it would be interesting to find an analytical derivation of the transformation formulae. One could envision a recursive strategy as in  the derivation of
formulae for overlaps between Bethe states, as implemented most recently for the general case of $\mathfrak{gl}(M|N)$ spin chains~\cite{Hutsalyuk:2017way}.  Alternatively, one
could envision a constructive proof determining the super determinant of the Gaudin matrix as the unique quantity transforming covariantly under fermionic duality.

\subsection*{Acknowledgements}

We would like to thank V.~Kazakov, C. Marboe and D.~Volin for interesting discussions. The figures were prepared using the \texttt{JaxoDraw} package \cite{Binosi:2003yf,Binosi:2008ig}.
The work of CK and DM was supported by DFF-FNU through grant number DFF-4002-00037. The work of KZ was supported 
by the grant "Exact Results in Gauge and String Theories" from the Knut and Alice Wallenberg foundation and by RFBR grant 18-01-00460 A. 

\appendix

\bibliographystyle{nb}

\begin{thebibliography}{10}
\providecommand{\href}[2]{#2}
\providecommand{\arxivref}[2]{\href{http://arxiv.org/abs/#1}{#2}}
\providecommand{\doiref}[2]{\href{http://dx.doi.org/#1}{#2}}
\providecommand{\nbbstauthor}[1]{#1}
\providecommand{\nbbstjournal}[1]{\textsf{#1}}
\providecommand{\nbbsttitle}[1]{\textit{#1}}
\providecommand{\nbbsturl}[1]{\texttt{#1}}
\providecommand{\nbbsteprint}[1]{\texttt{#1}}
\providecommand{\nbbststyle}{\raggedright\small\parskip0pt}
\nbbststyle

\bibitem{deLeeuw:2015hxa}
\nbbstauthor{M.~de~Leeuw, C.~Kristjansen and K.~Zarembo},
\nbbsttitle{``{One-point Functions in Defect CFT and Integrability}''},
\nbbstjournal{\doiref{10.1007/JHEP08(2015)098}{JHEP~1508,~098~(2015)}},
\nbbsteprint{\arxivref{1506.06958}{arxiv:1506.06958}}.

\bibitem{Buhl-Mortensen:2015gfd}
\nbbstauthor{I.~Buhl-Mortensen, M.~de~Leeuw, C.~Kristjansen and K.~Zarembo},
\nbbsttitle{``{One-point Functions in AdS/dCFT from Matrix Product States}''},
\nbbstjournal{\doiref{10.1007/JHEP02(2016)052}{JHEP~1602,~052~(2016)}},
\nbbsteprint{\arxivref{1512.02532}{arxiv:1512.02532}}.

\bibitem{Komatsu:2020sup}
\nbbstauthor{S.~Komatsu and Y.~Wang},
\nbbsttitle{``{Non-perturbative defect one-point functions in planar
  $\mathcal{N}=4$ super-Yang-Mills}''},
\nbbstjournal{\doiref{10.1016/j.nuclphysb.2020.115120}{Nucl.~Phys.~B~958,~115120~(2020)}},
\nbbsteprint{\arxivref{2004.09514}{arxiv:2004.09514}}.

\bibitem{Gombor:2020kgu}
\nbbstauthor{T.~Gombor and Z.~Bajnok},
\nbbsttitle{``{Boundary states, overlaps, nesting and bootstrapping
  AdS/dCFT}''},
\nbbstjournal{\doiref{10.1007/JHEP10(2020)123}{JHEP~2010,~123~(2020)}},
\nbbsteprint{\arxivref{2004.11329}{arxiv:2004.11329}}.

\bibitem{Kristjansen:2020mhn}
\nbbstauthor{C.~Kristjansen, D.~M\"uller and K.~Zarembo},
\nbbsttitle{``{Integrable boundary states in D3-D5 dCFT: beyond scalars}''},
\nbbstjournal{\doiref{10.1007/JHEP08(2020)103}{JHEP~2008,~103~(2020)}},
\nbbsteprint{\arxivref{2005.01392}{arxiv:2005.01392}}.

\bibitem{Gombor:2020auk}
\nbbstauthor{T.~Gombor and Z.~Bajnok},
\nbbsttitle{``{Boundary state bootstrap and asymptotic overlaps in
  AdS/dCFT}''},
\nbbsteprint{\arxivref{2006.16151}{arxiv:2006.16151}}.

\bibitem{Piroli:2017sei}
\nbbstauthor{L.~Piroli, B.~Pozsgay and E.~Vernier},
\nbbsttitle{``{What is an integrable quench?}''},
\nbbstjournal{\doiref{10.1016/j.nuclphysb.2017.10.012}{Nucl.~Phys.~B~925,~362~(2017)}},
\nbbsteprint{\arxivref{1709.04796}{arxiv:1709.04796}}.

\bibitem{Pozsgay:2009}
\nbbstauthor{B.~Pozsgay},
\nbbsttitle{``{Overlaps between eigenstates of the XXZ spin-1/2 chain and a
  class of simple product states}''},
\nbbstjournal{JSTAT~6,~{06011}~(2014)},
\nbbsteprint{\arxivref{1309.4593}{arxiv:1309.4593}}.

\bibitem{Brockmann:2014a}
\nbbstauthor{M.~Brockmann, J.~De~Nardis, B.~Wouters and J.-S.~Caux},
\nbbsttitle{``{A Gaudin-like determinant for overlaps of N\'{e}el and XXZ Bethe
  States}''},
\nbbstjournal{J.~Phys.~A:~Math.~Theor.~47,~145003~(2014)},
\nbbsteprint{\arxivref{1401.2877}{arxiv:1401.2877}}.

\bibitem{Brockmann:2014b}
\nbbstauthor{M.~Brockmann, J.~De~Nardis, B.~Wouters and J.-S.~Caux},
\nbbsttitle{``{N\'{e}el-XXZ state overlaps: odd particle numbers and
  Lieb-Liniger scaling limit}''},
\nbbstjournal{J.~Phys.~A:~Math.~Theor.~47,~345003~(2014)},
\nbbsteprint{\arxivref{1403.7469}{arxiv:1403.7469}}.

\bibitem{Pozsgay:2018ixm}
\nbbstauthor{B.~Pozsgay},
\nbbsttitle{``{Overlaps with arbitrary two-site states in the XXZ spin
  chain}''},
\nbbstjournal{\doiref{10.1088/1742-5468/aabbe1}{J.~Stat.~Mech.~1805,~053103~(2018)}},
\nbbsteprint{\arxivref{1801.03838}{arxiv:1801.03838}}.

\bibitem{deLeeuw:2016umh}
\nbbstauthor{M.~de~Leeuw, C.~Kristjansen and S.~Mori},
\nbbsttitle{``{AdS/dCFT one-point functions of the SU(3) sector}''},
\nbbstjournal{\doiref{10.1016/j.physletb.2016.10.044}{Phys.~Lett.~B~763,~197~(2016)}},
\nbbsteprint{\arxivref{1607.03123}{arxiv:1607.03123}}.

\bibitem{deLeeuw:2018mkd}
\nbbstauthor{M.~De~Leeuw, C.~Kristjansen and G.~Linardopoulos},
\nbbsttitle{``{Scalar one-point functions and matrix product states of
  AdS/dCFT}''},
\nbbstjournal{\doiref{10.1016/j.physletb.2018.03.083}{Phys.~Lett.~B~781,~238~(2018)}},
\nbbsteprint{\arxivref{1802.01598}{arxiv:1802.01598}}.

\bibitem{Piroli:2018ksf}
\nbbstauthor{L.~Piroli, E.~Vernier, P.~Calabrese and B.~Pozsgay},
\nbbsttitle{``{Integrable quenches in nested spin chains I: the exact steady
  states}''},
\nbbstjournal{\doiref{10.1088/1742-5468/ab1c51}{J.~Stat.~Mech.~1906,~063103~(2019)}},
\nbbsteprint{\arxivref{1811.00432}{arxiv:1811.00432}}.

\bibitem{Piroli:2018don}
\nbbstauthor{L.~Piroli, E.~Vernier, P.~Calabrese and B.~Pozsgay},
\nbbsttitle{``{Integrable quenches in nested spin chains II: fusion of boundary
  transfer matrices}''},
\nbbstjournal{\doiref{10.1088/1742-5468/ab1c52}{J.~Stat.~Mech.~1906,~063104~(2019)}},
\nbbsteprint{\arxivref{1812.05330}{arxiv:1812.05330}}.

\bibitem{deLeeuw:2019ebw}
\nbbstauthor{M.~De~Leeuw, T.~Gombor, C.~Kristjansen, G.~Linardopoulos and
  B.~Pozsgay},
\nbbsttitle{``{Spin Chain Overlaps and the Twisted Yangian}''},
\nbbstjournal{\doiref{10.1007/JHEP01(2020)176}{JHEP~2001,~176~(2020)}},
\nbbsteprint{\arxivref{1912.09338}{arxiv:1912.09338}}.

\bibitem{Gaudin:1976sv}
\nbbstauthor{M.~Gaudin},
\nbbsttitle{``{Diagonalisation d'une Classe d'Hamiltoniens de Spin}''},
\nbbstjournal{J.~Phys.~France~37,~1087~(1976)}.

\bibitem{Korepin:1982gg}
\nbbstauthor{V.~Korepin},
\nbbsttitle{``{Calculation of norms of Bethe wave functions}''},
\nbbstjournal{\doiref{10.1007/BF01212176}{Commun.~Math.~Phys.~86,~391~(1982)}}.

\bibitem{Frappat:1996pb}
\nbbstauthor{L.~Frappat, P.~Sorba and A.~Sciarrino},
\nbbsttitle{``{Dictionary on Lie superalgebras}''},
\nbbsteprint{\arxivref{hep-th/9607161}{hep-th/9607161}}.



\bibitem{Tsuboi:1998ne}
\nbbstauthor{Z.~Tsuboi},
\nbbsttitle{``{Analytic Bethe Ansatz And Functional Equations Associated With
  Any Simple Root Systems Of The Lie Superalgebra $sl(r+1|s+1)$}''},
\nbbstjournal{Physica~A252,~565~(1998)}.

\bibitem{Gromov:2017blm}
\nbbstauthor{N.~Gromov},
\nbbsttitle{``{Introduction to the Spectrum of $N=4$ SYM and the Quantum
  Spectral Curve}''},
\nbbsteprint{\arxivref{1708.03648}{arxiv:1708.03648}}.


\bibitem{Beisert:2005fw}
\nbbstauthor{N.~Beisert and M.~Staudacher},
\nbbsttitle{``Long-range $PSU(2,2|4)$ Bethe ansaetze for gauge theory and
  strings''},
\nbbstjournal{Nucl.~Phys.~B727,~1~(2005)},
\nbbsteprint{\arxivref{hep-th/0504190}{hep-th/0504190}}.

\bibitem{Beisert:2003jj}
\nbbstauthor{N.~Beisert},
\nbbsttitle{``{The complete one-loop dilatation operator of N = 4 super
  Yang-Mills theory}''},
\nbbstjournal{\doiref{10.1016/j.nuclphysb.2003.10.019}{Nucl.~Phys.~B676,~3~(2004)}},
\nbbsteprint{\arxivref{hep-th/0307015}{hep-th/0307015}}.

\bibitem{Jiang:2019xdz}
\nbbstauthor{Y.~Jiang, S.~Komatsu and E.~Vescovi},
\nbbsttitle{``{Structure constants in $ \mathcal{N} $ = 4 SYM at finite
  coupling as worldsheet g-function}''},
\nbbstjournal{\doiref{10.1007/JHEP07(2020)037}{JHEP~2007,~037~(2020)}},
\nbbsteprint{\arxivref{1906.07733}{arxiv:1906.07733}}.

\bibitem{Jiang:2019zig}
\nbbstauthor{Y.~Jiang, S.~Komatsu and E.~Vescovi},
\nbbsttitle{``{Exact Three-Point Functions of Determinant Operators in Planar
  $N=4$ Supersymmetric Yang-Mills Theory}''},
\nbbstjournal{\doiref{10.1103/PhysRevLett.123.191601}{Phys.~Rev.~Lett.~123,~191601~(2019)}},
\nbbsteprint{\arxivref{1907.11242}{arxiv:1907.11242}}.

\bibitem{Gaiotto:2008sa}
\nbbstauthor{D.~Gaiotto and E.~Witten},
\nbbsttitle{``{Supersymmetric Boundary Conditions in N=4 Super Yang-Mills
  Theory}''},
\nbbstjournal{\doiref{10.1007/s10955-009-9687-3}{J.~Statist.~Phys.~135,~789~(2009)}},
\nbbsteprint{\arxivref{0804.2902}{arxiv:0804.2902}}.

\bibitem{Buhl-Mortensen:2016pxs}
\nbbstauthor{I.~Buhl-Mortensen, M.~de~Leeuw, A.~C.~Ipsen, C.~Kristjansen and
  M.~Wilhelm},
\nbbsttitle{``{One-loop one-point functions in gauge-gravity dualities with
  defects}''},
\nbbstjournal{\doiref{10.1103/PhysRevLett.117.231603}{Phys.~Rev.~Lett.~117,~231603~(2016)}},
\nbbsteprint{\arxivref{1606.01886}{arxiv:1606.01886}}.

\bibitem{Buhl-Mortensen:2016jqo}
\nbbstauthor{I.~Buhl-Mortensen, M.~de~Leeuw, A.~C.~Ipsen, C.~Kristjansen and
  M.~Wilhelm},
\nbbsttitle{``{A Quantum Check of AdS/dCFT}''},
\nbbstjournal{\doiref{10.1007/JHEP01(2017)098}{JHEP~1701,~098~(2017)}},
\nbbsteprint{\arxivref{1611.04603}{arxiv:1611.04603}}.

\bibitem{Buhl-Mortensen:2017ind}
\nbbstauthor{I.~Buhl-Mortensen, M.~de~Leeuw, A.~C.~Ipsen, C.~Kristjansen and
  M.~Wilhelm},
\nbbsttitle{``{Asymptotic One-Point Functions in Gauge-String Duality with
  Defects}''},
\nbbstjournal{\doiref{10.1103/PhysRevLett.119.261604}{Phys.~Rev.~Lett.~119,~261604~(2017)}},
\nbbsteprint{\arxivref{1704.07386}{arxiv:1704.07386}}.

\bibitem{Ghoshal:1993tm}
\nbbstauthor{S.~Ghoshal and A.~B.~Zamolodchikov},
\nbbsttitle{``{Boundary S matrix and boundary state in two-dimensional
  integrable quantum field theory}''},
\nbbstjournal{\doiref{10.1142/S0217751X94001552}{Int.~J.~Mod.~Phys.~A9,~3841~(1994)}},
\nbbsteprint{\arxivref{hep-th/9306002}{hep-th/9306002}},
[Erratum: Int. J. Mod. Phys.A9,4353(1994)].

\bibitem{Beisert:2005di}
\nbbstauthor{N.~Beisert, V.~A.~Kazakov, K.~Sakai and K.~Zarembo},
\nbbsttitle{``{Complete spectrum of long operators in N = 4 SYM at one
  loop}''},
\nbbstjournal{\doiref{10.1088/1126-6708/2005/07/030}{JHEP~0507,~030~(2005)}},
\nbbsteprint{\arxivref{hep-th/0503200}{hep-th/0503200}}.

\bibitem{Kazakov:2007fy}
\nbbstauthor{V.~Kazakov, A.~Sorin and A.~Zabrodin},
\nbbsttitle{``{Supersymmetric Bethe ansatz and Baxter equations from discrete
  Hirota dynamics}''},
\nbbstjournal{\doiref{10.1016/j.nuclphysb.2007.06.025}{Nucl.~Phys.~B790,~345~(2008)}},
\nbbsteprint{\arxivref{hep-th/0703147}{hep-th/0703147}}.

\bibitem{Gromov:2014caa}
\nbbstauthor{N.~Gromov, V.~Kazakov, S.~Leurent and D.~Volin},
\nbbsttitle{``{Quantum spectral curve for arbitrary state/operator in
  AdS$_{5}$/CFT$_{4}$}''},
\nbbstjournal{\doiref{10.1007/JHEP09(2015)187}{JHEP~1509,~187~(2015)}},
\nbbsteprint{\arxivref{1405.4857}{arxiv:1405.4857}}.


\bibitem{Tsuboi:2009ud}
\nbbstauthor{Z.~Tsuboi},
\nbbsttitle{``{Solutions of the T-system and Baxter equations for
  supersymmetric spin chains}''},
\nbbstjournal{\doiref{10.1016/j.nuclphysb.2009.08.009}{Nucl.~Phys.~B~826,~399~(2010)}},
\nbbsteprint{\arxivref{0906.2039}{arxiv:0906.2039}}.



\bibitem{Gromov:2010km}
\nbbstauthor{N.~Gromov, V.~Kazakov, S.~Leurent and Z.~Tsuboi},
\nbbsttitle{``{Wronskian Solution for AdS/CFT Y-system}''},
\nbbstjournal{\doiref{10.1007/JHEP01(2011)155}{JHEP~1101,~155~(2011)}},
\nbbsteprint{\arxivref{1010.2720}{arxiv:1010.2720}}.

\bibitem{Kazakov:2010iu}
\nbbstauthor{V.~Kazakov, S.~Leurent and Z.~Tsuboi},
\nbbsttitle{``{Baxter's Q-operators and operatorial Backlund flow for quantum
  (super)-spin chains}''},
\nbbstjournal{\doiref{10.1007/s00220-012-1428-9}{Commun.~Math.~Phys.~311,~787~(2012)}},
\nbbsteprint{\arxivref{1010.4022}{arxiv:1010.4022}}.

\bibitem{Bazhanov:2010jq}
\nbbstauthor{V.~V.~Bazhanov, R.~Frassek, T.~Lukowski, C.~Meneghelli and
  M.~Staudacher},
\nbbsttitle{``{Baxter Q-Operators and Representations of Yangians}''},
\nbbstjournal{\doiref{10.1016/j.nuclphysb.2011.04.006}{Nucl.~Phys.~B850,~148~(2011)}},
\nbbsteprint{\arxivref{1010.3699}{arxiv:1010.3699}}.

\bibitem{Frassek:2010ga}
\nbbstauthor{R.~Frassek, T.~Lukowski, C.~Meneghelli and M.~Staudacher},
\nbbsttitle{``{Oscillator Construction of $su(n|m)$ Q-Operators}''},
\nbbstjournal{\doiref{10.1016/j.nuclphysb.2011.04.008}{Nucl.~Phys.~B~850,~175~(2011)}},
\nbbsteprint{\arxivref{1012.6021}{arxiv:1012.6021}}.

\bibitem{Kazakov:2015efa}
\nbbstauthor{V.~Kazakov, S.~Leurent and D.~Volin},
\nbbsttitle{``{T-system on T-hook: Grassmannian Solution and Twisted Quantum
  Spectral Curve}''},
\nbbstjournal{\doiref{10.1007/JHEP12(2016)044}{JHEP~1612,~044~(2016)}},
\nbbsteprint{\arxivref{1510.02100}{arxiv:1510.02100}}.

\bibitem{Marboe:2016yyn}
\nbbstauthor{C.~Marboe and D.~Volin},
\nbbsttitle{``{Fast analytic solver of rational Bethe equations}''},
\nbbstjournal{\doiref{10.1088/1751-8121/aa6b88}{J.~Phys.~A~50,~204002~(2017)}},
\nbbsteprint{\arxivref{1608.06504}{arxiv:1608.06504}}.

\bibitem{Kazakov:2018ugh}
\nbbstauthor{V.~Kazakov},
\nbbsttitle{``{Quantum Spectral Curve of $\gamma$-twisted ${\cal N}=4$ SYM
  theory and fishnet CFT}''},
\nbbstjournal{\doiref{10.1142/9789813233867_0016}{Rev.~Math.~Phys.~30,~1840010~(2018)}},
\nbbsteprint{\arxivref{1802.02160}{arxiv:1802.02160}}.

\bibitem{deLeeuw:2017dkd}
\nbbstauthor{M.~de~Leeuw, A.~C.~Ipsen, C.~Kristjansen, K.~E.~Vardinghus and
  M.~Wilhelm},
\nbbsttitle{``{Two-point functions in AdS/dCFT and the boundary conformal
  bootstrap equations}''},
\nbbstjournal{\doiref{10.1007/JHEP08(2017)020}{JHEP~1708,~020~(2017)}},
\nbbsteprint{\arxivref{1705.03898}{arxiv:1705.03898}}.

\bibitem{Beisert:2003xu}
\nbbstauthor{N.~Beisert, J.~A.~Minahan, M.~Staudacher and K.~Zarembo},
\nbbsttitle{``{Stringing spins and spinning strings}''},
\nbbstjournal{\doiref{10.1088/1126-6708/2003/09/010}{JHEP~0309,~010~(2003)}},
\nbbsteprint{\arxivref{hep-th/0306139}{hep-th/0306139}}.

\bibitem{Nepomechie:2013mua}
\nbbstauthor{R.~I.~Nepomechie and C.~Wang},
\nbbsttitle{``{Algebraic Bethe ansatz for singular solutions}''},
\nbbstjournal{\doiref{10.1088/1751-8113/46/32/325002}{J.~Phys.~A~46,~325002~(2013)}},
\nbbsteprint{\arxivref{1304.7978}{arxiv:1304.7978}}.

\bibitem{Beisert:2003yb}
\nbbstauthor{N.~Beisert and M.~Staudacher},
\nbbsttitle{``The {$\mathcal{N}=\mathord{}$4} SYM Integrable Super Spin
  Chain''},
\nbbstjournal{Nucl.~Phys.~B670,~439~(2003)},
\nbbsteprint{\arxivref{hep-th/0307042}{hep-th/0307042}}.

\bibitem{inprogress}
\nbbstauthor{C.~Kristjansen, D.~M\"uller and K.~Zarembo},
\nbbsttitle{``{Work in progress}''}.

\bibitem{Hutsalyuk:2017way}
\nbbstauthor{A.~Hutsalyuk, A.~Liashyk, S.~Pakuliak, E.~Ragoucy and N.~Slavnov},
\nbbsttitle{``{Norm of Bethe vectors in models with $gl(m|n)$ symmetry}''},
\nbbstjournal{\doiref{10.1016/j.nuclphysb.2017.11.006}{Nucl.~Phys.~B~926,~256~(2018)}},
\nbbsteprint{\arxivref{1705.09219}{arxiv:1705.09219}}.

\bibitem{Binosi:2003yf}
\nbbstauthor{D.~Binosi and L.~Theussl},
\nbbsttitle{``{JaxoDraw: A Graphical user interface for drawing Feynman
  diagrams}''},
\nbbstjournal{\doiref{10.1016/j.cpc.2004.05.001}{Comput.~Phys.~Commun.~161,~76~(2004)}},
\nbbsteprint{\arxivref{hep-ph/0309015}{hep-ph/0309015}}.

\bibitem{Binosi:2008ig}
\nbbstauthor{D.~Binosi, J.~Collins, C.~Kaufhold and L.~Theussl},
\nbbsttitle{``{JaxoDraw: A Graphical user interface for drawing Feynman
  diagrams. Version 2.0 release notes}''},
\nbbstjournal{\doiref{10.1016/j.cpc.2009.02.020}{Comput.~Phys.~Commun.~180,~1709~(2009)}},
\nbbsteprint{\arxivref{0811.4113}{arxiv:0811.4113}}.

\end{thebibliography}


\end{document}